\documentclass[11pt,a4paper]{article}
\usepackage[utf8]{inputenc}

\usepackage{jheppub}
\usepackage{amsmath}
\usepackage{breqn}
\usepackage{comment}
\usepackage[colorinlistoftodos]{todonotes}
\usepackage{float}
\usepackage{graphics}

\graphicspath{{Figure/}}
\usepackage{caption} 
\usepackage{subcaption} 
\usepackage{amsfonts}
\usepackage{todonotes} 

\title{Two-loop helicity amplitudes for diphoton production with massive quark loop} 

\author{Taushif Ahmed$^{a}$, Amlan Chakraborty$^{b,c}$, Ekta Chaubey$^{d}$ and Mandeep Kaur$^{e}$}
\emailAdd{taushif.ahmed@ur.de, amlan.chakraborty@unimi.it, eekta@uni-bonn.de, mandeep.kaur@desy.de}
\affiliation{
$^a$Institute for Theoretical Physics, University of Regensburg, 93040 Regensburg, Germany 
\\$^b$Tif Lab, Dipartimento di Fisica, Universit\'{a} di Milano, Sezione di Milano, Via Celoria 16, I-20133 Milano, Italy
\\$^c$The Institute of Mathematical Sciences, 600113 Chennai, India
\\$^d$Bethe Center for Theoretical Physics, Universitaet Bonn, 53115 Bonn, Germany
\\$^e$Deutsches Elektronen-Synchrotron DESY,
Platanenallee 6, 15738 Zeuthen, Germany}

\preprint{ TIF-UNIMI-2025-4, BONN-TH-2025-01, DESY-25-017}

\abstract{
We compute two-loop helicity amplitudes in QCD for diphoton production through quark- and gluon-initiated channels, accounting for a massive internal quark loop by keeping its full mass dependence. Using physical projectors, we directly decompose the amplitude into its helicity components. By renormalising the heavy quark mass in on-shell, and other quantities in $\overline{\rm MS}$ schemes, we obtain finite remainders. This work paves the way for calculating the cross-section for diphoton production at higher orders in QCD with a massive quark loop, employing different subtraction schemes. The effect of a heavy quark is expected to play a crucial role in high-luminosity LHC.
}

\newcommand{\as}{\alpha_s}

\newcommand{\asb}{\alpha_{s,b}}

\newcommand{\ep}{\epsilon}
\newcommand{\slashed}{\slash \hspace{-0.19cm}}

\newcommand{\be}{\begin{equation}}
\newcommand{\ee}{\end{equation}}
\newcommand{\bea}{\begin{eqnarray}}
\newcommand{\eea}{\end{eqnarray}}

\definecolor{Red}{rgb}{1.,0.,0.}
\definecolor{randomcolour}{rgb}{0.2,0.5,0.7}

\DeclareMathAlphabet\mathbfcal{OMS}{cmsy}{b}{n}

\def\OMIT#1{}

\definecolor{darkred}{rgb}{0.9,0,0}

\definecolor{darkgreen}{rgb}{0,0,0.9}

\definecolor{darkblue}{rgb}{0,0,0.9}

\usepackage{cancel}
\allowdisplaybreaks[1]

\begin{document}
\allowdisplaybreaks[4]
\unitlength1cm
\maketitle
\flushbottom


\def\D{{\cal D}}
\def\DD{\overline{\cal D}}
\def\g{\overline{\cal C}}
\def\gm{\gamma}
\def\M{{\cal M}}
\def\ep{\epsilon}
\def\epm1{\frac{1}{\epsilon}}
\def\epm2{\frac{1}{\epsilon^{2}}}
\def\epm3{\frac{1}{\epsilon^{3}}}
\def\epm4{\frac{1}{\epsilon^{4}}}
\def\unM{\hat{\cal M}}
\def\ashat{\hat{a}_{s}}
\def\asmur{a_{s}^{2}(\mu_{R}^{2})}
\def\sigbar{{{\overline {\sigma}}}\left(a_{s}(\mu_{R}^{2}), L\left(\mu_{R}^{2}, m_{H}^{2}\right)\right)}
\def\sigbarn{{{{\overline \sigma}}_{n}\left(a_{s}(\mu_{R}^{2}) L\left(\mu_{R}^{2}, m_{H}^{2}\right)\right)}}
\def\unas{ \left( \frac{\hat{a}_s}{\mu_0^{\epsilon}} S_{\epsilon} \right) }
\def\rnM{{\cal M}}
\def\bt{\beta}
\def\cD{{\cal D}}
\def\cC{{\cal C}}
\def\ca{\text{\tiny C}_\text{\tiny A}}
\def\cf{\text{\tiny C}_\text{\tiny F}}
\def\ct{{\red []}}
\def\sv{\text{SV}}
\def\murOmu{\left( \frac{\mu_{R}^{2}}{\mu^{2}} \right)}
\def\bb{b{\bar{b}}}
\def\bt0{\beta_{0}}
\def\bt1{\beta_{1}}
\def\bt2{\beta_{2}}
\def\bt3{\beta_{3}}
\def\gm0{\gamma_{0}}
\def\gm1{\gamma_{1}}
\def\gm2{\gamma_{2}}
\def\gm3{\gamma_{3}}
\def\nn{\nonumber}
\def\l{\left}
\def\r{\right}
\def\T{{\cal Z}}    
\def\U{{\cal Y}}

\def\nn{\nonumber\\}
\def\ep{\epsilon}
\def\T{\mathcal{T}}
\def\V{\mathcal{V}}

\def\qgraf{{\fontfamily{qcr}\selectfont
QGRAF}}
\def\python{{\fontfamily{qcr}\selectfont
PYTHON}}
\def\form{{\fontfamily{qcr}\selectfont
FORM}}
\def\reduze{{\fontfamily{qcr}\selectfont
REDUZE2}}
\def\kira{{\fontfamily{qcr}\selectfont
Kira}}
\def\litered{{\fontfamily{qcr}\selectfont
LiteRed}}
\def\fire{{\fontfamily{qcr}\selectfont
FIRE5}}
\def\air{{\fontfamily{qcr}\selectfont
AIR}}
\def\mint{{\fontfamily{qcr}\selectfont
Mint}}
\def\hepforge{{\fontfamily{qcr}\selectfont
HepForge}}
\def\arXiv{{\fontfamily{qcr}\selectfont
arXiv}}
\def\Python{{\fontfamily{qcr}\selectfont
Python}}
\def\ginac{{\fontfamily{qcr}\selectfont
GiNaC}}
\def\polylogtools{{\fontfamily{qcr}\selectfont
PolyLogTools}}
\def\anci{{\fontfamily{qcr}\selectfont
Finite\_ppbk.m}}

\newcommand{\dis}{}
\newcommand{\overbar}[1]{mkern-1.5mu\overline{\mkern-1.5mu#1\mkern-1.5mu}\mkern
1.5mu}

\section{Introduction}
\label{sec:intro} 

The production of diphotons ($\gamma\gamma$) at high-energy colliders, such as the Large Hadron Collider (LHC), serves as an important process in probing the Standard Model (SM) and exploring potential new physics~\cite{
CMS:2016kgr,ATLAS:2017ayi}. Diphoton final states provide a clean experimental signature due to the excellent photon identification and reconstruction capabilities in modern detectors. They also play a crucial role in precision studies, such as measuring Higgs boson properties, testing perturbative Quantum Chromodynamics (QCD), and searching for exotic particles or phenomena. The differential cross-section of this process has been precisely measured at both the Tevatron~\cite{CDF:2012ool, D0:2013gxo} and the LHC~\cite{CMS:2011xtn, CMS:2014mvm}. This signature was pivotal as one of the two “golden channels” that led to the discovery of the Higgs boson~\cite{ATLAS:2012yve, CMS:2012qbp}. The $H \to \gamma\gamma$ decay remains one of the cleanest final states for exploring the properties of the Higgs boson and its production mechanisms.

At hadron colliders, diphoton production at leading order (LO) originates from the annihilation of a quark and an antiquark via the process $q\bar{q} \to \gamma\gamma$. Corrections at next-to-LO (NLO) in the strong coupling constant ($\alpha_S$) for this process were computed decades ago in ref.~\cite{Binoth:1999qq}. Subsequent developments have extended this to next-to-NLO (NNLO) accuracy ($\mathcal{O}(\alpha_S^2)$)~\cite{Catani:2011qz,Campbell:2016yrh,Catani:2018krb,Schuermann:2022qdm}, with results implemented in public computational tools such as $2\gamma$\textsc{NNLO}~\cite{Catani:2011qz}, {\sc MCFM}~\cite{Campbell:2016yrh}, and {\sc Matrix}~\cite{Grazzini:2017mhc}. The relevant scattering amplitudes in massless QCD have been extensively studied in refs.~\cite{Dicus:1987fk,DelDuca:1999pa,Anastasiou:2002zn,DelDuca:2003uz}. Currently, it is available up to three loops~\cite{Caola:2020dfu}. In refs.~\cite{Chawdhry:2020for,Agarwal:2021grm,Chawdhry:2021mkw,Agarwal:2021vdh}, the two-loop amplitude associated with a jet has been computed. These form a building block for next-to-NNLO (N$^3$LO) corrections in massless QCD. 
The inclusion of massive quark in the loop starts appearing only at NNLO $(\alpha_S^2)$ level, as shown in figure~\ref{fig:process1-figs}. In ref.~\cite{Campbell:2016yrh}, the effect of the top quark was discussed. Recently, the full phenomenological study has been conducted in ref.~\cite{Becchetti:2023yat} and the underlying two-loop helicity amplitudes have been presented in ref.~\cite{Becchetti:2023wev} where the master integrals were evaluated employing generalised power series method.

At NNLO, a new production channel emerges: the fusion of gluons into a diphoton pair, mediated by a quark loop, as shown in figure~\ref{fig:ggaa-qqaa-1L} for a massive quark. The gluon-induced contribution is not only finite and gauge-invariant on its own but also unusually significant due to the large gluon-gluon luminosity at hadron colliders. Its contribution is of the size of born subprocess $q\bar{q} \to \gamma\gamma$.
Higher-order corrections to this gluon fusion channel, specifically at $\mathcal{O}(\alpha_S^3)$, involve two-loop contributions to $gg \to \gamma\gamma$. The two-loop computation for the massless QCD case was first carried out in ref.~\cite{Bern:2001df}, while ref.~\cite{Chawdhry:2021hkp} extended this work to include configurations involving an associated jet. Their phenomenological analysis in massless QCD was performed in~\cite{Campbell:2016yrh,Bern:2002jx}. Currently, the amplitude is available at three-loop order~\cite{Bargiela:2021wuy}.
Although the fully analytic two-loop amplitude for this process, including a top quark loop, has not yet been presented in the literature, its impact on the cross-section has been explored. Previous studies have relied on numerical~\cite{Maltoni:2018zvp} and semi-numerical~\cite{Chen:2019fla} evaluations, with the latter incorporating analytic expressions for the subset of master integrals that were available at the time.

The \textit{goal} of this article is to present, for the first time, the computation of the two-loop amplitude for $gg \to \gamma\gamma$, retaining the full top-quark mass dependence within the loop and expressing the result in terms of analytic functions. In addition, we also compute the two-loop amplitude for $q\bar q \to \gamma \gamma$ with the massive top quark in the loop. The quark-initiated two-loop amplitude contributes at NNLO ($\mathcal{O}(\alpha_s^2)$), while the gluon-initiated counterpart appears at N$^3$LO ($\mathcal{O}(\alpha_s^3)$) in QCD at hadron colliders.

Representing helicity amplitudes in analytic form is not only essential for advancing our understanding of quantum field theory but also ensures numerical stability in cross-section computations and other related observables.  Furthermore, it will be valuable to evaluate cross-sections using various subtraction schemes; in particular, studying results derived from a local subtraction framework would be interesting. This virtual amplitude is one of the most important ingredients for this endeavour. This will also pave the way to compute the helicity amplitudes for dijet production at the LHC by incorporating the mass of top quark in the loop. 
From a technical perspective, an important aspect of this work involves investigating the impact of the elliptic sector that arises from one of the non-planar integral families, providing deeper insights into the structure of these amplitudes; the numerical evaluation of elliptic integrals is still not as efficient as other non-elliptic Feynman integrals.

We adopt the method of projecting the amplitude onto the helicity basis using physical projectors, as described in refs.~\cite{Peraro:2019cjj,Peraro:2020sfm}. An alternative approach to constructing physical projectors is discussed in ref.~\cite{Chen:2019wyb}. The bare integrand is generated and processed through a series of in-house codes implemented in \texttt{FORM}~\cite{Vermaseren:2000nd}. The associated Feynman integrals are subsequently processed through \texttt{Kira}~\cite{Maierhofer:2017gsa,Klappert:2020nbg} to apply integration-by-parts identities (IBP)~\cite{Chetyrkin:1979bj,Chetyrkin:1981qh} to express the integrand in terms of a minimal set of master integrals. These integrals have been extensively studied in the literature~\cite{Caron-Huot:2014lda,Becchetti:2017abb,AH:2023ewe,Becchetti:2023wev,Becchetti:2023yat}. In ref.~\cite{Ahmed:2024tsg}, the final missing set of master integrals containing elliptic sectors was evaluated by some of us, thereby enabling the complete analytic computation of these amplitudes. 
While many of these integrals exist in various forms in the literature, we independently set up a comprehensive system of differential equations containing all uncrossed master integrals to ensure a consistent representation.
The bare helicity amplitudes are renormalised in a mixed scheme: we adopt the on-shell scheme for mass renormalisation, while the remaining quantities are renormalised in the $\overline{\rm MS}$. We provide the helicity amplitudes in terms of a set of master integrals as an ancillary file~\cite{zenodokaur25}. The finite remainder is
available upon request from the authors for those interested. We present a few benchmark numerical values of all helicity amplitudes, in particular, around the top quark threshold.

The article is organised as follows. Section~\ref{sec:kin} describes the kinematic setup of the process including its Lorentz covariant decomposition. In section~\ref{sec:hel}, we describe the method of constructing helicity amplitudes and the procedure to get the bare integrand. The ultraviolet renormalisation and infrared factorisation are discussed in section~\ref{sec:subtraction}. In section~\ref{sec:res}, we discuss the results and their numerical implementation. We also describe the checks performed to ensure the correctness of the results. We conclude with our outlook in section~\ref{sec:conc}.

\section{Setup}
\label{sec:kin}
We consider the following scattering processes:
\begin{equation} \label{eq:processes}
\begin{split}
& g (p_1) + g(p_2)  + \gamma(p_3) + \gamma(p_4) \to 0 \,, \\
& q (p_1) + \bar{q}(p_2) + \gamma(p_3) + \gamma(p_4) \to 0 \,.
\end{split}
\end{equation}
We label the momenta of the particles by $p_1,\cdots, p_4$ and regard all of them as incoming that satisfy
\begin{equation}
p_1+p_2+p_3+p_4=0\,, \qquad p_i^2=0 .
\end{equation}
The physical di-photon production at the LHC can be obtained from \eqref{eq:processes} by crossing $p_{3,4} \to -p_{3,4}$. In computing an observable, such as cross-section for the di-photon production, one requires $g \gamma \to g \gamma$ and $q \gamma  \to q \gamma$ channels which can also be obtained by crossing from \eqref{eq:processes}.
The kinematic Mandelstam invariants of the process,
\begin{equation}
s = (p_1+ p_2)^2\,,~ t = (p_2+p_3)^2\,,~u = (p_1+p_3)^2\,,
\end{equation}
are related by momentum conservation $s+t+u = 0$. Consequently, no Euclidean region exists kinematically for the scattering process, rendering it interesting to study. The $2 \to 2$ physical region corresponds to the scattering region
\begin{equation}
\label{eq:anCont}
s>0\,, t<0\,, u<0\,.
\end{equation}
We construct two dimensionless parameters as
\begin{align}
    x=\frac{s}{m_t^2}, \quad y=\frac{t}{m_t^2},
\end{align}
where $m_t$ denotes the mass of the top quark.

\begin{figure}[h!]
\centering
\includegraphics[width = 1.2in]{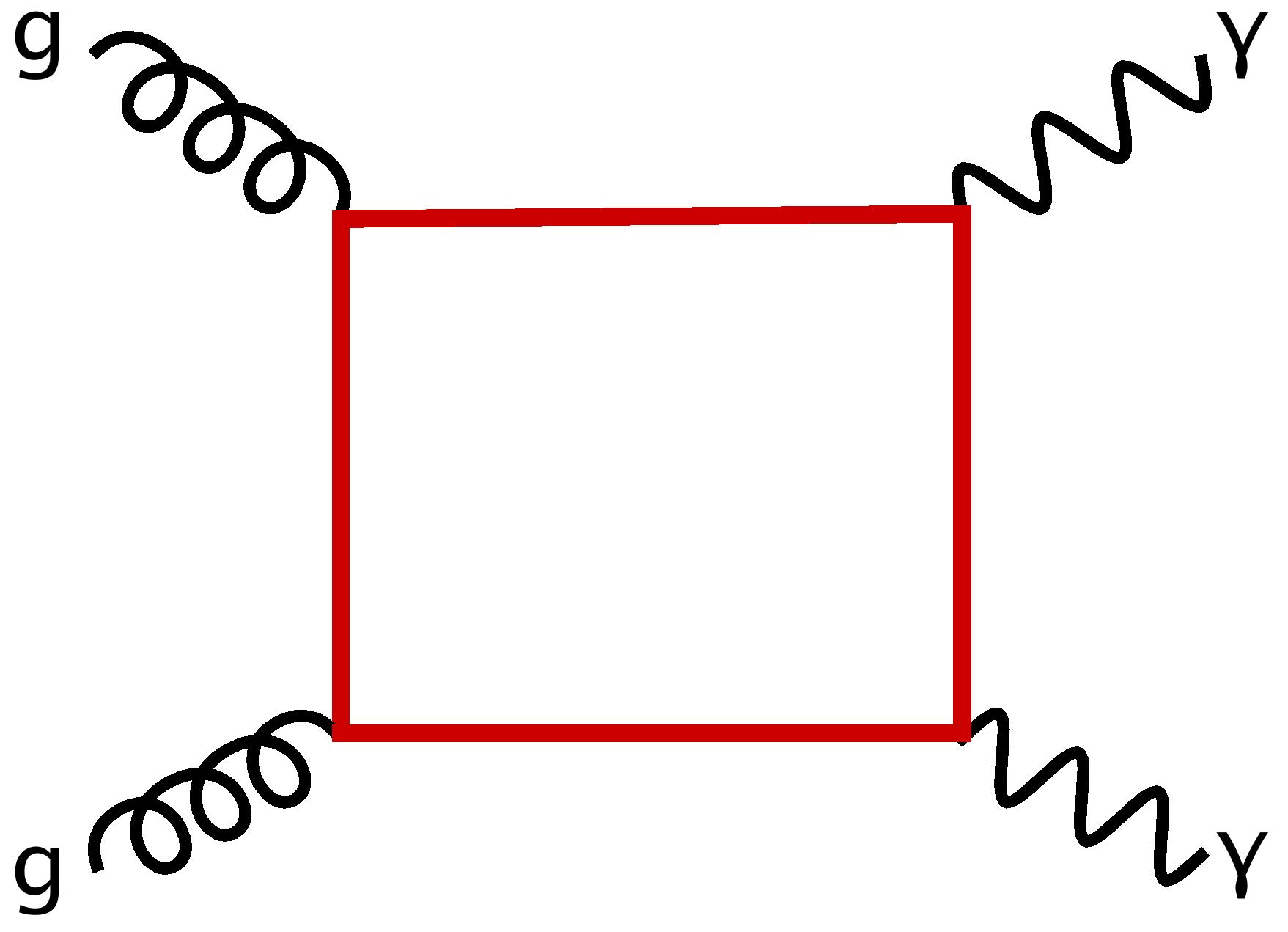}
\caption{Representative leading order Feynman diagrams for gluon-initiated channel involving massive quark loop. The red lines represent top-quark, while the curly and wavy black lines correspond to gluon and photon, respectively.}
\label{fig:ggaa-qqaa-1L}
\end{figure}

In this article, we consider the scattering with at least one massive quark in the loop. So, both are loop-induced processes, as shown in figure~\ref{fig:ggaa-qqaa-1L} and \ref{fig:process1-figs}.
Our goal is to calculate the two-loop amplitude of these processes in QCD. We denote the mass of the massive quark by $m_t$. 
The amplitude can be rewritten by factoring out the overall color factor as
\begin{align}
\mathcal{A} = \mathcal{C} \, A \, ,
\end{align}
where 
\begin{align}\label{eq:color}  
\mathcal{C}  =
    \begin{cases}
         \delta^{a_1 a_2}, & \text{for } g g \to \gamma \gamma, \\[8pt]
        \delta^{i_1 i_2}, & \text{for } q \bar{q} \to \gamma \gamma.
    \end{cases}
\end{align}
Here, $i_n(a_n)$ represents an $SU(N_c)$ index in the fundamental (adjoint) representation. The partial amplitude $A$ depends on the number of active massless ($n_f$) and massive ($n_{f_t}$) quark flavors, as well as their respective electric charges, denoted by $Q_f$ and $Q_{f_t}$. Since we focus on Feynman diagrams that include at least one massive quark loop, meaning the lowest power of $n_{f_t}$ contributing to the amplitude is 1. After extracting all color structures, the partial amplitude can further be decomposed into a basis of $N_T$ independent Lorentz covariant tensor structures $T_i$ as
\begin{equation}
A = \sum_{i=1}^{N_T} \mathcal{F}_i T_i  \,,
\label{eq:ampoT}
\end{equation}
where $\mathcal{F}_i$ are called the form factors. These form factors can be expanded perturbatively in powers of the strong coupling constant, $\alpha_s$.

We adopt the ’t Hooft–Veltman (tHV) regularisation scheme~\cite{tHooft:1972tcz}, in which loop momenta are treated in $d = 4 - 2\epsilon$ dimensions, while external momenta and polarisations remain in four dimensions. Within this framework, we follow the method proposed in~\cite{Peraro:2019cjj,Peraro:2020sfm}, which eliminates the evanescent $(-2\epsilon)$-dimensional helicity states and allows us to work with a set of tensors $T_i$ whose number corresponds directly to the independent helicity configurations. A similar approach can be found in refs.~\cite{Chen:2019wyb,Ahmed:2019udm}. 
 
In $g g \to \gamma \gamma$ channel, there are $N_T = 8$ independent tensor structures. By adopting the cyclic gauge choice, $\epsilon_i \cdot p_{i+1} = 0$ (with $p_5 \equiv p_1$), and applying the transversality condition, $\epsilon_i \cdot p_i = 0$, we obtain the following results~\cite{Peraro:2019cjj,Peraro:2020sfm,Caola:2021izf}:
\begin{align}
\label{eq:tengg}
T_1^{g} &= p_1\!\cdot\!\ep_2 \; p_1\!\cdot\!\ep_3 \; p_2\!\cdot\!\ep_4 \; p_3\!\cdot\!\ep_1, \quad
T_2^{g} = \ep_3\!\cdot\!\ep_4 \; p_1\!\cdot\!\ep_2 \; p_3\!\cdot\!\ep_1 , \quad
T_3^{g} = \ep_2\!\cdot\!\ep_4 \; p_1\!\cdot\!\ep_3 \; p_3\!\cdot\!\ep_1 , \notag\\
T_4^{g} &= \ep_2\!\cdot\!\ep_3 \; p_2\!\cdot\!\ep_4 \; p_3\!\cdot\!\ep_1 , \quad
T_5^{g} = \ep_1\!\cdot\!\ep_4 \; p_1\!\cdot\!\ep_2 \; p_1\!\cdot\!\ep_3 , \quad
T_6^{g} = \ep_1\!\cdot\!\ep_3 \; p_1\!\cdot\!\ep_2 \; p_2\!\cdot\!\ep_4 , \notag \\
T_7^{g} &= \ep_1\!\cdot\!\ep_2 \; p_1\!\cdot\!\ep_3 \; p_2\!\cdot\!\ep_4 , \quad
T_8^{g} = \ep_1\!\cdot\!\ep_2 \; \ep_3\!\cdot\!\ep_4 + \ep_1\!\cdot\!\ep_4 \; \ep_2\!\cdot\!\ep_3 + \ep_1\!\cdot\!\ep_3 \; \ep_2\!\cdot\!\ep_4 .
\end{align}
The polarisation vector is denoted by $\ep(p_i) \equiv \ep_i$.
Unlike in tHV scheme, in conventional dimensional regularisation, one requires 10 tensorial structures~\cite{Binoth:2002xg,Ahmed:2019qtg}.
In $q \bar{q} \to \gamma \gamma$ channel, $N_t=4$ and with the gauge choice $\ep_3 \cdot p_2 = \ep_4 \cdot p_1 = 0$, we get~\cite{Caola:2022dfa}
\begin{align}
\label{eq:tenqq}
T_1^q &= \bar{u}(p_2) \slashed{\ep_3} u(p_1) \: \ep_4 \!\cdot\! p_2, \quad
T_2^q = \bar{u}(p_2) \slashed{\ep_3} u(p_1) \: \ep_4 \!\cdot\! p_1, \notag\\
T_3^q &= \bar{u}(p_2) \slashed{p_3} u(p_1) \: \ep_3 \!\cdot\! p_1 \, \ep_4 \!\cdot\! p_2, \quad
T_4^q = \bar{u}(p_2) \slashed{p_3} u(p_1) \: \ep_3 \!\cdot\! \ep_4\, .
\end{align}
The form factors $\mathcal{F}_i$ can be extracted from $A$ with appropriate projectors $P_j$, defined to satisfy the orthogonality condition $\sum_{pol} P_j T_i^{g|q} = \delta_{ji}$. The superscript $g|q$ denotes either gluon- or quark-initiated channel.

\section{Helicity Amplitudes}
\label{sec:hel}

To compute the helicity amplitudes $A_{\vec{\lambda}}$, it suffices to evaluate the tensors $T_i$ for specific helicity configurations $\vec{\lambda} = \{\lambda_1, \lambda_2, \lambda_3, \lambda_4\}$ of the external particles. Each helicity amplitude corresponding to a given configuration $\vec{\lambda}$ can then be expressed as a linear combination of the form factors $\mathcal{F}_i$ as 
\begin{equation}
A_{\vec{\lambda}}
= \sum_{i=1}^{N_T} T_{i,{\vec{\lambda}}} \mathcal{F}_i
= \mathcal{S}_{\vec{\lambda}} \, \mathcal{H}_{\vec{\lambda}}\,.
\end{equation}
The overall spinor factors $\mathcal{S}_{\vec{\lambda}}$ can be extracted from $A_{\vec{\lambda}}$ using the spinor-helicity formalism. For a detailed introduction to this approach, we refer to ref.~\cite{Dixon:1996wi}. In this formalism, external quarks with fixed helicities are defined as
\begin{equation}
| p \rangle = \overline{[ p |} = \frac{1+\gamma_5}{2} u(p)\,, \quad
| p ] = \overline{\langle p |} = \frac{1-\gamma_5}{2} u(p)\,,
\end{equation}
with $[ p | = \overline{u}(p) \frac{1-\gamma_5}{2}$ and
$\langle p | = u(p) \frac{1+\gamma_5}{2}$
treating particles and anti-particles on an equal footing,
while polarisation vectors take the following form
\begin{equation}
\epsilon^\mu_{j,+} = \frac{\langle p_j | \gamma^\mu | q_j ] }{ \sqrt{2} [ p_j q_j ]}\,, \quad
\epsilon^\mu_{j,-} = \frac{\langle q_j | \gamma^\mu | p_j ] }{ \sqrt{2} \langle q_j p_j \rangle }\,,
\end{equation}
where $q_i$ is the massless reference vector corresponding to the $i$-th external gluon and is chosen consistently with the gauge conditions used to determine the tensor bases of eqs.~\eqref{eq:tengg} and~\eqref{eq:tenqq}.  
For the $g g \to \gamma \gamma$ channel, there are 8 independent helicity amplitudes which are related to the remaining ones through parity as
\begin{align}
    A^g_{\vec{\lambda}}=A^g_{-\vec{\lambda}}(\langle ij \rangle \leftrightarrow [ji]).
\end{align}
Here the negative sign flips the helicity. We choose independent $\vec{\lambda}=\{++++,-+++,+-++,++-+,+++-,--++,-+-+,+--+\}$. By choosing the reference vector $q_i = p_{i+1}$, where we identify $p_5 \equiv p_1$, we have the following spinor factors~\cite{Bern:2001df}
\begin{align}
\mathcal{S}_{++++}^g &= \frac{\langle 1 2 \rangle \langle 3 4 \rangle}{[1 2] [3 4]} \,, &
\mathcal{S}_{-+++}^g &= \frac{[1 2][1 4]\langle 2 4 \rangle}{[3 4][2 3][2 4]} \,, &
\mathcal{S}_{+-++}^g &= \frac{[2 1][2 4]\langle 1 4 \rangle}{[3 4][1 3][1 4]} \,, \notag \\
\mathcal{S}_{++-+}^g &= \frac{[3 2][3 4]\langle 2 4 \rangle}{[1 4][2 1][2 4]} \,, &
\mathcal{S}_{+++-}^g &= \frac{[4 2][4 3]\langle 2 3 \rangle}{[1 3][2 1][2 3]} \,, &
\mathcal{S}_{--++}^g &= \frac{[1 2]\langle 3 4 \rangle}{\langle 1 2 \rangle [3 4]} \,, \notag \\
\mathcal{S}_{-+-+}^g &= \frac{[1 3]\langle 2 4 \rangle}{\langle 1 3 \rangle [2 4]} \,, &
\mathcal{S}_{+--+}^g &= \frac{[2 3]\langle 1 4 \rangle}{\langle 2 3 \rangle [1 4]} \,. &&
\label{eq:helamp}
\end{align}
For the $q \bar{q} \to \gamma \gamma$ channel, we have 4 independent helicity amplitudes which can be used to obtain the remaining 4 through charge-conjugation as
\begin{align}
    A^q_{+-\lambda_3\lambda_4}=A^q_{-+\lambda_3^*\lambda_4^*}(\langle ij \rangle \leftrightarrow [ji])\,.
\end{align}
The $\lambda^*$ refers opposite helicity of $\lambda$. We choose $q_3 = p_2, \; q_4=p_1$ and define the spinor factors as
\begin{align}
\mathcal{S}_{-+--}^q &= \frac{2[3 4]^2}{\langle1 3\rangle[2 3]} \,, & 
\mathcal{S}_{-+-+}^q &= \frac{2\langle2 4\rangle[1 3]}{\langle2 3\rangle[2 4]} \,,
\notag\\
\mathcal{S}_{-++-}^q &= \frac{2\langle2 3\rangle[4 1]}{\langle2 4\rangle[3 2]} \,,&
\mathcal{S}_{-+++}^q &= \frac{2\langle3 4\rangle^2}{\langle3 1\rangle[2 3]} \,.
\label{eq:spin_weight_qqga}
\end{align}
In our conventions, all external legs are treated as incoming. For outgoing particles, the helicities of the respective legs must be reversed. The spinor inner products are defined as $\langle ij \rangle = \langle i^- | j^+ \rangle$ and $[ij] = \langle i^+ | j^- \rangle$, where $\lvert i^\pm \rangle$ represent massless Weyl spinors associated with the momentum $p_i$ and labeled by their helicity sign. These inner products are antisymmetric and have magnitudes given by $|\langle ij \rangle| = |[ij]| = \sqrt{s_{ij}}$, where $s_{ij} = 2p_i \cdot p_j$ are the usual Mandelstam invariants: $s_{12}=s,\; s_{23}=t,\; s_{13}=u$. Consequently, the helicity-dependent factors $S_{\lambda_1 \lambda_2 \lambda_3 \lambda_4}^{g|q}$, derived from these spinor products, are pure phases.

The spinor-free helicity amplitude $\mathcal{H}_{\vec{\lambda}}$ can be expanded in powers of bare strong coupling $\asb$ as
\begin{align} \label{eq:helicity_pert}
\mathcal{H}_{\vec{\lambda}}^{g|q} &= 
4 \pi \alpha
\sum_{\ell=0}^{2} \left(\frac{\asb}{4\pi}\right)^{\ell} \mathcal{H}^{^{g|q},(\ell)}_{\vec{\lambda}} + \mathcal{O}(\alpha_{s,b}^3)\,,
\end{align}
where we factor out an overall term proportional to the square of the electric charge, $e^2 = 4\pi\alpha$. The quantity $\mathcal{H}^{^{g|q},(\ell)}_{\vec{\lambda}}$ represents the bare $\ell$-loop amplitude.  
It is important to note that, as the $g g \to \gamma \gamma$ channel is loop-induced, the leading-order term in its perturbative expansion vanishes. In contrast, the $q \bar{q} \to \gamma \gamma$ channel contributes non-trivially to all three orders.  
For the quark-initiated processes involving at least one massive quark loop, non-zero diagrams begin to appear only at the two-loop level. However, through renormalization, the lower-order diagrams also contribute indirectly to the overall result. 
\begin{figure}[ht!]
\centering
{\includegraphics[width = 1.1in]{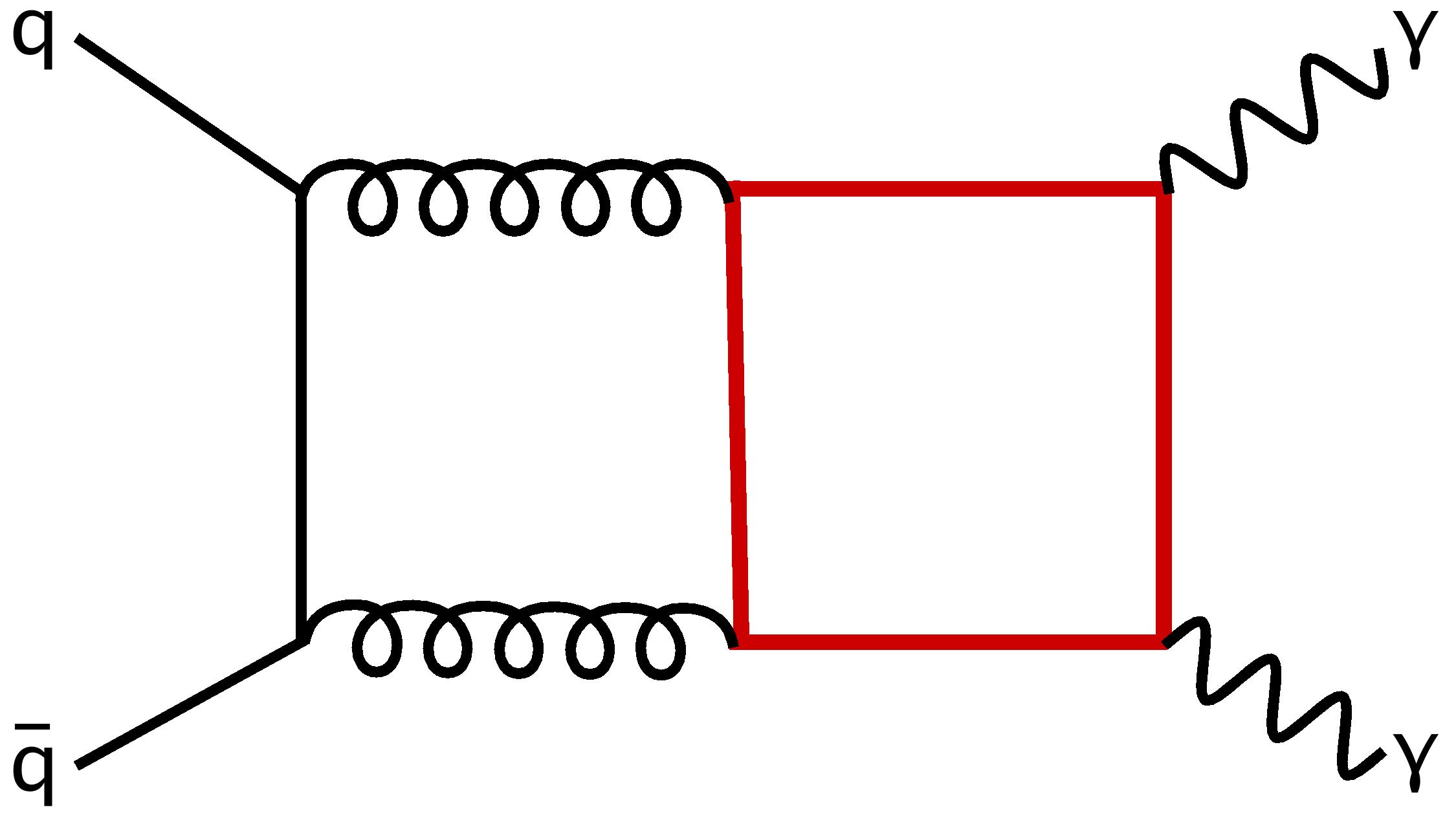}} 
\hspace{1.5cm}
{\includegraphics[width = 1.1in]{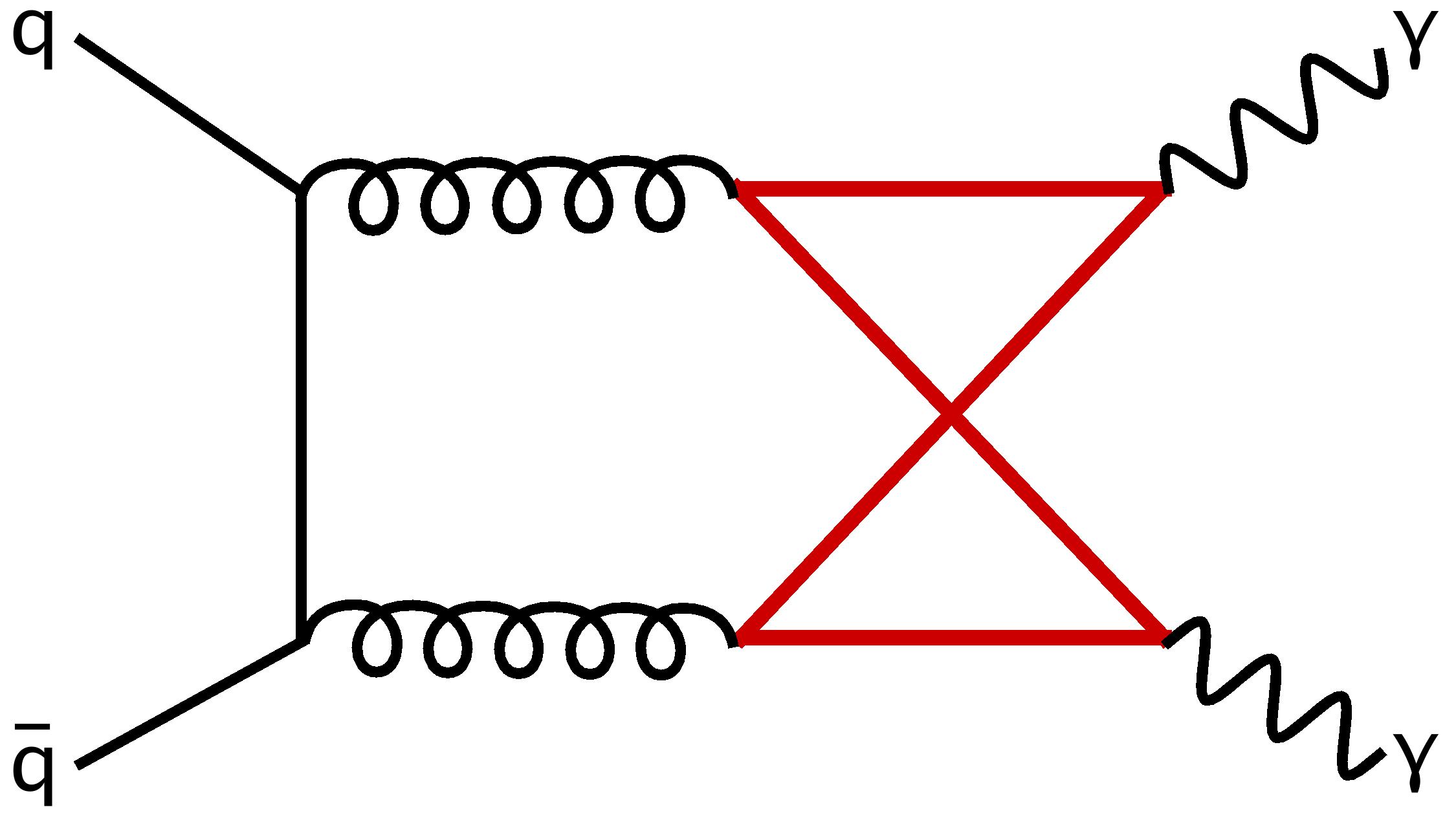}}
\hspace{1.5cm}
{\includegraphics[width = 0.9in]{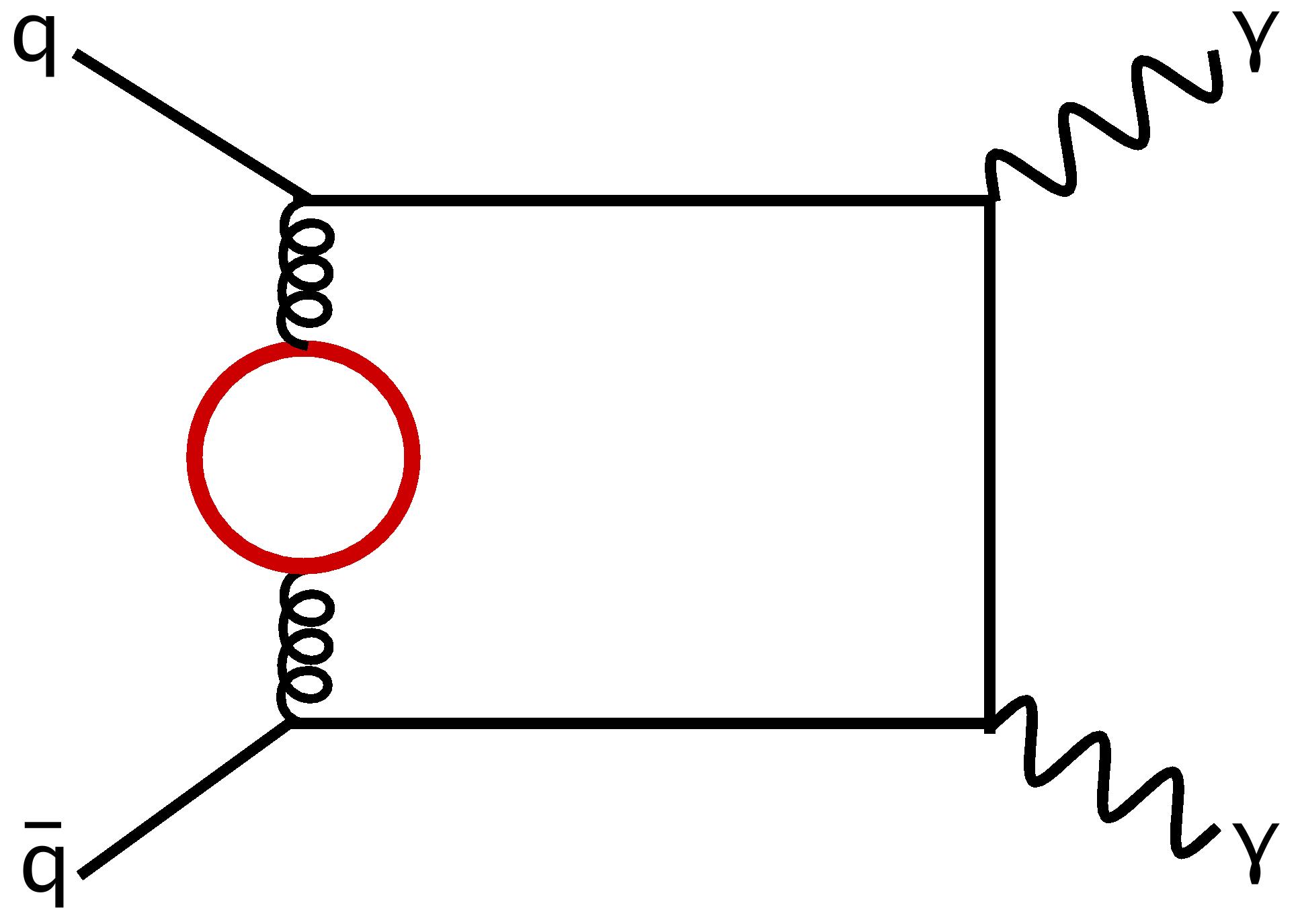}}
\caption{{{Representative two-loop Feynman diagrams for $q\bar{q}$ channel. }}}
\label{fig:process1-figs}
\end{figure}
\begin{figure}[t!]
\centering
{\includegraphics[width = 1.1in]{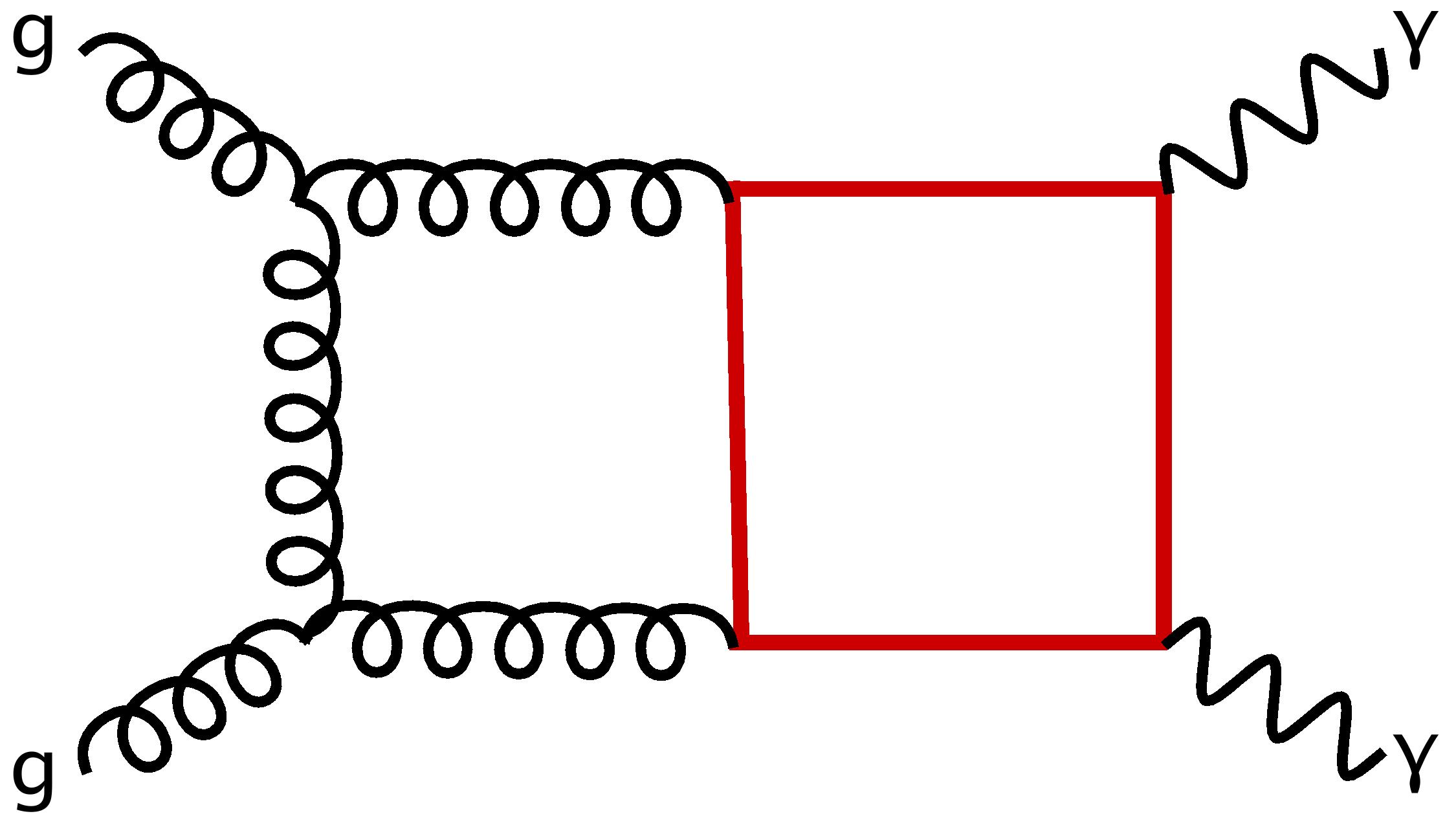}} 
\hspace{0.5cm}
{\includegraphics[width = 1.1in]{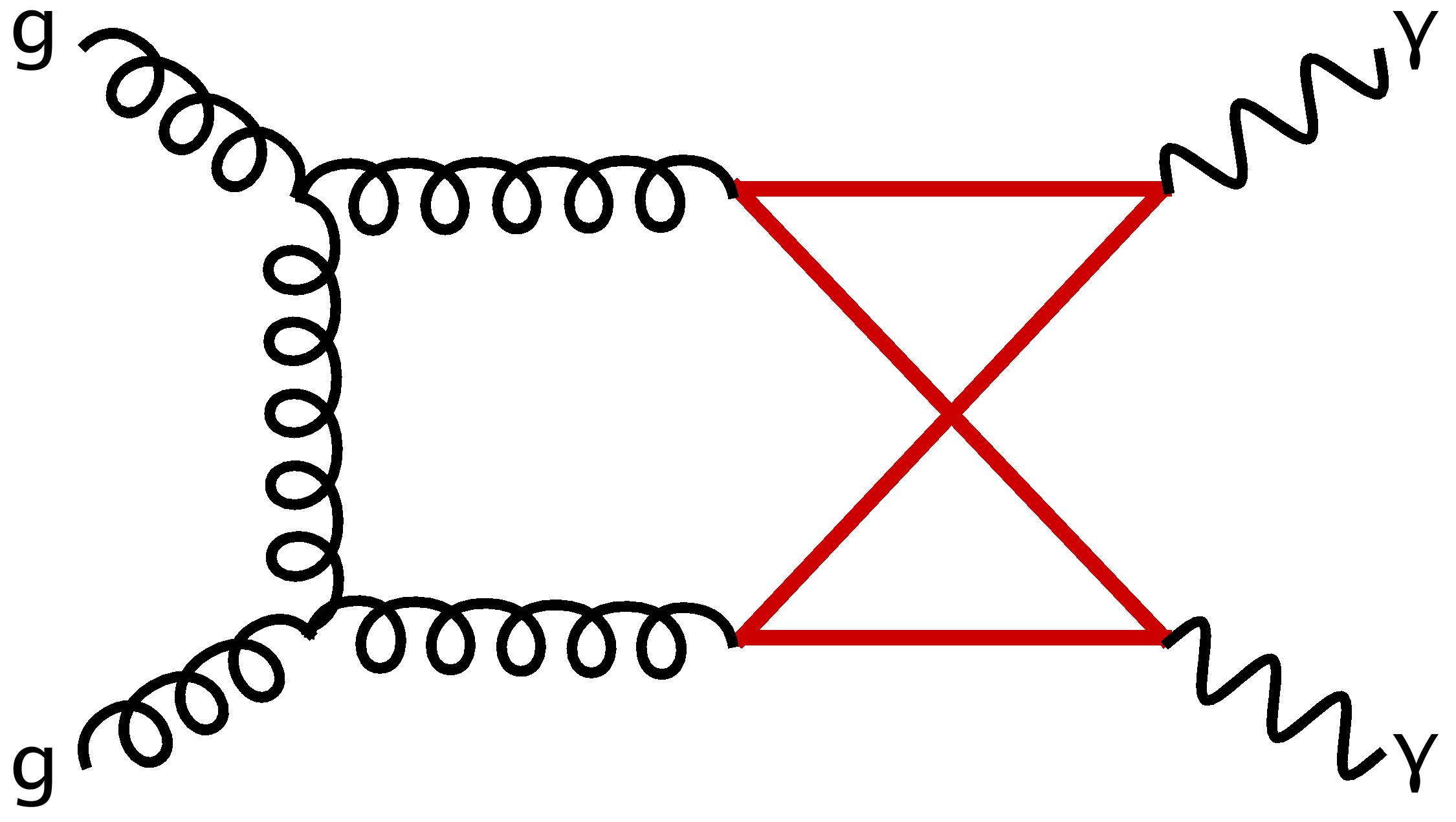}}
\hspace{0.5cm}
{\includegraphics[width = 1.1in]{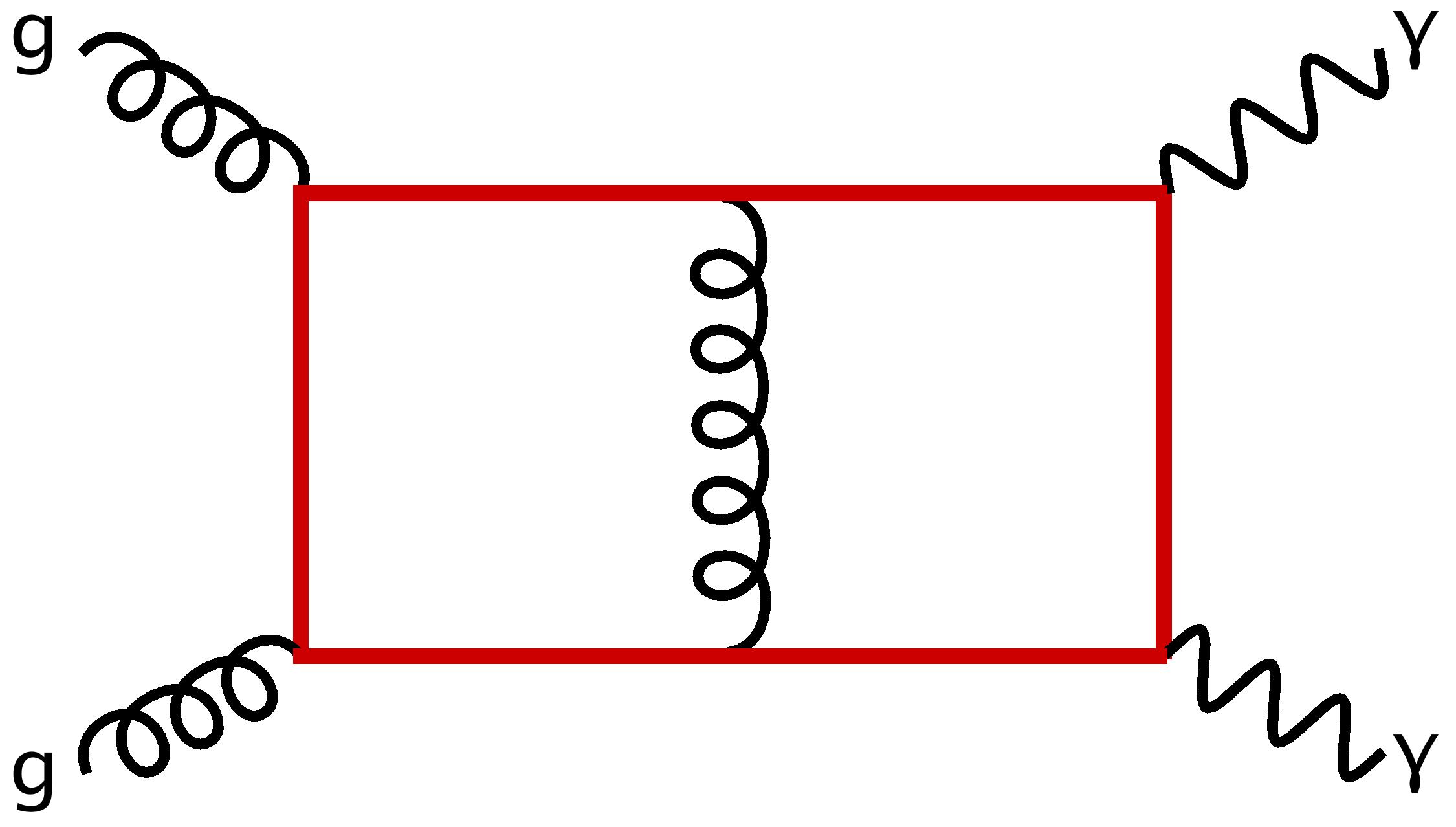}}
\hspace{0.5cm}
{\includegraphics[width = 1.1in]{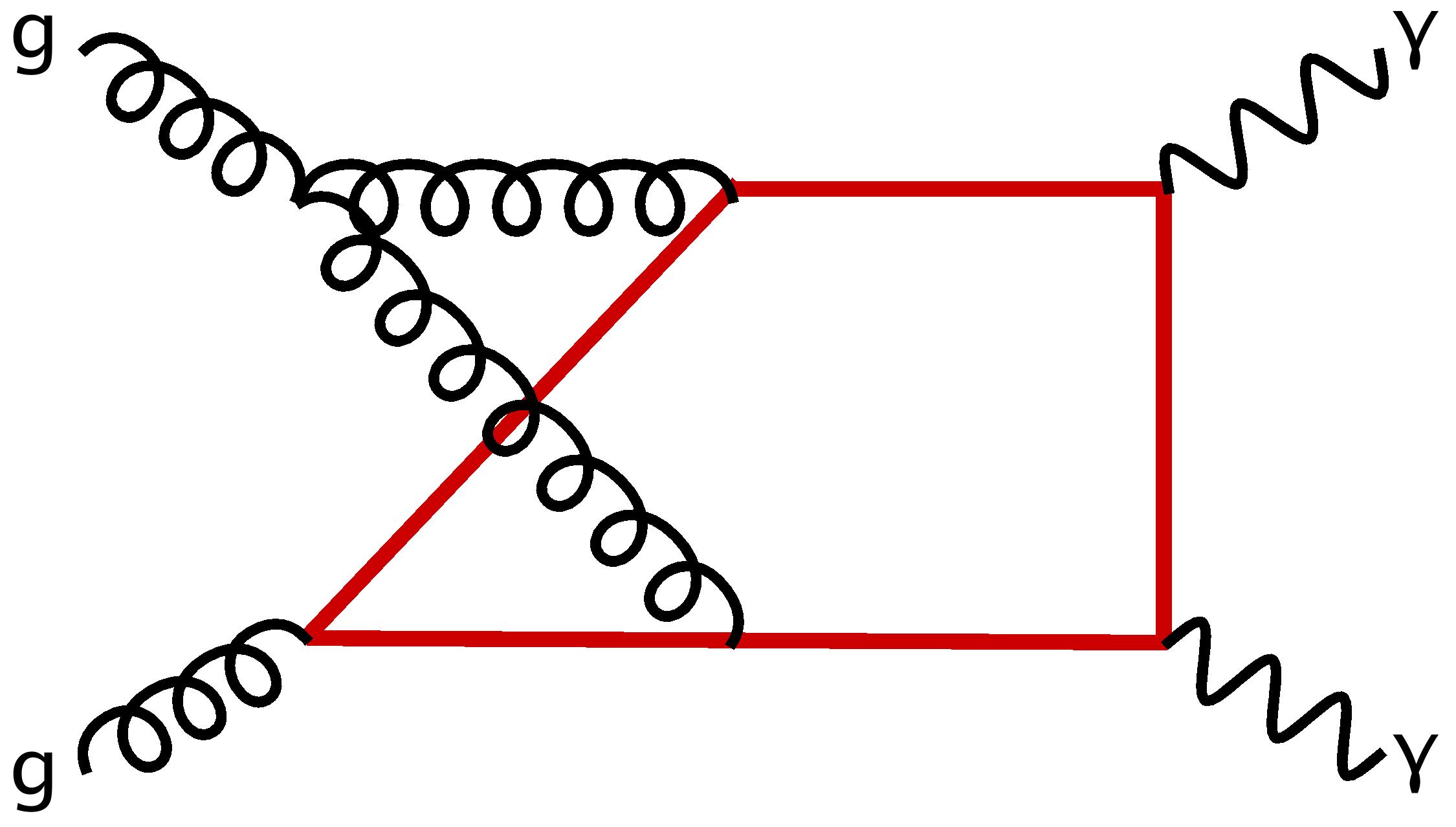}}
\caption{{{Representative two-loop Feynman diagrams for $g g$ channel. }}}
\label{fig:process2-figs}
\end{figure}

We generate the Feynman diagrams for each channel using \texttt{Qgraf}\cite{Nogueira:1991ex}. There are 8 diagrams at one-loop for $gg$ channel. At two loops, the $g g$ channel comprises 166 diagrams, while the $q \bar{q}$ channel contains 55 diagrams. Samples of the two-loop diagrams are illustrated in figures~\ref{fig:process1-figs} and~\ref{fig:process2-figs}.
To process these diagrams, we use \texttt{FORM}~\cite{Vermaseren:2000nd}, applying the tensor projectors defined in eqs.~\eqref{eq:tengg} and~\eqref{eq:tenqq}. We evaluate the Dirac traces and simplify the colour algebra using in-house codes. The latter involves repeated application of standard colour identities,
\begin{align}
  {({T^a})}_{ij} {({T^a})}_{kh} &= \frac12
  \left({\delta}_{ih} {\delta}_{kj} - \frac{1}{N_c} {\delta}_{ij} {\delta}_{kh}\right)\,, \quad
f^{abc} = - 2 \: i \: {\rm Tr}(T^a[T^b,T^c])\,.
\end{align}
The form factors are expressed as linear combinations of scalar Feynman integrals, with rational coefficients that depend on the Mandelstam invariants $s$, $t$, mass $m_t$, and the dimensional regulator $\epsilon$.
The form factors for the $g g \to \gamma \gamma$ process involve 26,577 scalar Feynman integrals, while the $q \bar{q} \to \gamma \gamma$ process requires 2,289 integrals.
We parametrize the $\ell$-loop Feynman integrals as follows:
\begin{equation}\label{integrals}
{I}^{\text{top}}_{n_1,n_2,...,n_N} = \mu_0^{2L\epsilon} e^{L \epsilon \gamma_E}  \int \prod_{i=1}^L \left( \frac{\mathrm{d}^d k_i}{i \pi^{\frac{d}{2}}} \right) \frac{1}{D_1^{n_1}D_2^{n_2} \dots D_N^{n_N}}.
\end{equation}
Here, $\gamma_E \approx 0.5772$ is the Euler-Mascheroni constant, and $\mu_0$ is the dimensional regularization scale. The factor $e^{L \epsilon \gamma_E}$ is purely conventional and is chosen for later convenience, while the factor $\mu_0^{2L\epsilon}$ ensures that the integrals maintain integer mass dimensions.
For a general process with $E$ independent external momenta and $L$ loops, one requires $L(L+1)/2 + L E$ independent denominators to describe all possible scalar products of loop momenta with either loop or external momenta. A specific complete set of denominators ${D_i}$ at a given loop order is typically referred to as an integral family.
We organize the amplitude into as few integral families as possible, allowing for permutations of external momenta (crossings). At two loops, this requires two planar and two non-planar families, which we present in tabular form in Table~\ref{table:1}.
\begin{table}[htb]
\raggedleft
\resizebox{\textwidth}{!}{%
\begin{tabular}{ |c| c|  c| c| c|}
\hline
 Family & PL1 & PL2 & NPL1 & NPL2 \\
\hline
$D_1  $& $k_1^2 - m_t^2$  & $k_1^2$ & $k_1^2$& $(k_1-p_1)^2$\\
$D_2 $& $(k_1+p_1)^2- m_t^2$ & $(k_1+p_1)^2$ & $(k_1+p_1)^2$ & $k_1^2$ \\  
$D_3 $& $(k_1+p_1+p_2)^2- m_t^2$ & $(k_1+p_1+p_2)^2$  & $(k_1 + k_2)^2-m_t^2$ & $(k_1+p_2)^2$\\
$D_4 $& $(k_1 + k_2)^2$ & $(k_1 + k_2)^2-m_t^2$ & $k_2^2- m_t^2$ & $(k_1+k_2-p_1)^2- m_t^2$\\
$D_5 $& $k_2^2- m_t^2$ & $k_2^2- m_t^2$ & $(k_2 + p_3)^2- m_t^2$ & $k_2^2- m_t^2$  \\
$D_6$& $(k_2 + p_3)^2- m_t^2$ & $(k_2 + p_3)^2- m_t^2$ & $(k_2 - p_1-p_2)^2- m_t^2$& $(k_2+p_3)^2- m_t^2$ \\
$D_7 $& $(k_2 - p_1-p_2)^2- m_t^2$ & $(k_2 - p_1-p_2)^2- m_t^2$ & $(k_1 + k_2-p_2)^2-m_t^2$& $(k_1+k_2+p_2+p_3)^2- m_t^2$ \\
$D_8$ & $(k_2 - p_1)^2- m_t^2$ & $(k_2 - p_1)^2- m_t^2$  & $(k_2 - p_1)^2- m_t^2$ & $(k_1 
+p_3)^2$ \\
$D_9$ & $(k_1 - p_3)^2- m_t^2$ & $(k_1 - p_3)^2$ & $(k_1 - p_3)^2$ & $(k_2 - p_1)^2-m_t^2$ \\
\hline
\end{tabular}%
}
\caption{Planar and non-planar integral families at two loops. The first seven entries denote the real propagators appearing in Feynman diagrams. All diagrams are mapped to these and their crossed families.}
\label{table:1}
\end{table}
There, we indicate the loop momenta with $k_1$ and $k_2$.  We name PL1 and PL2 the families corresponding to the planar graphs and NPL1,  NPL2 the ones corresponding to the non-planar graphs. 
\begin{figure}[t!]
\centering
\includegraphics[width = 1.1in]{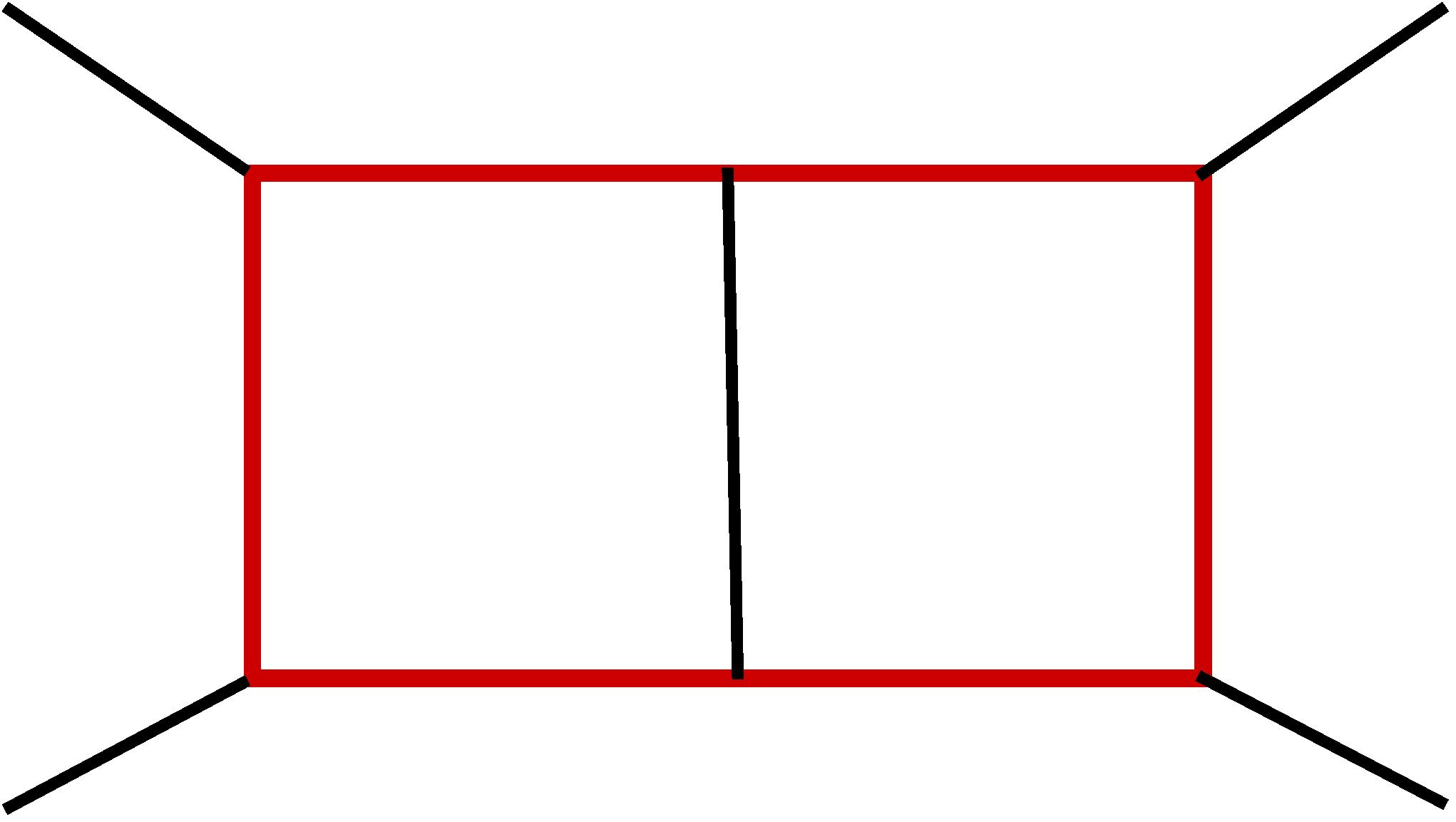}
\hspace{0.5cm}
\includegraphics[width = 1.1in]{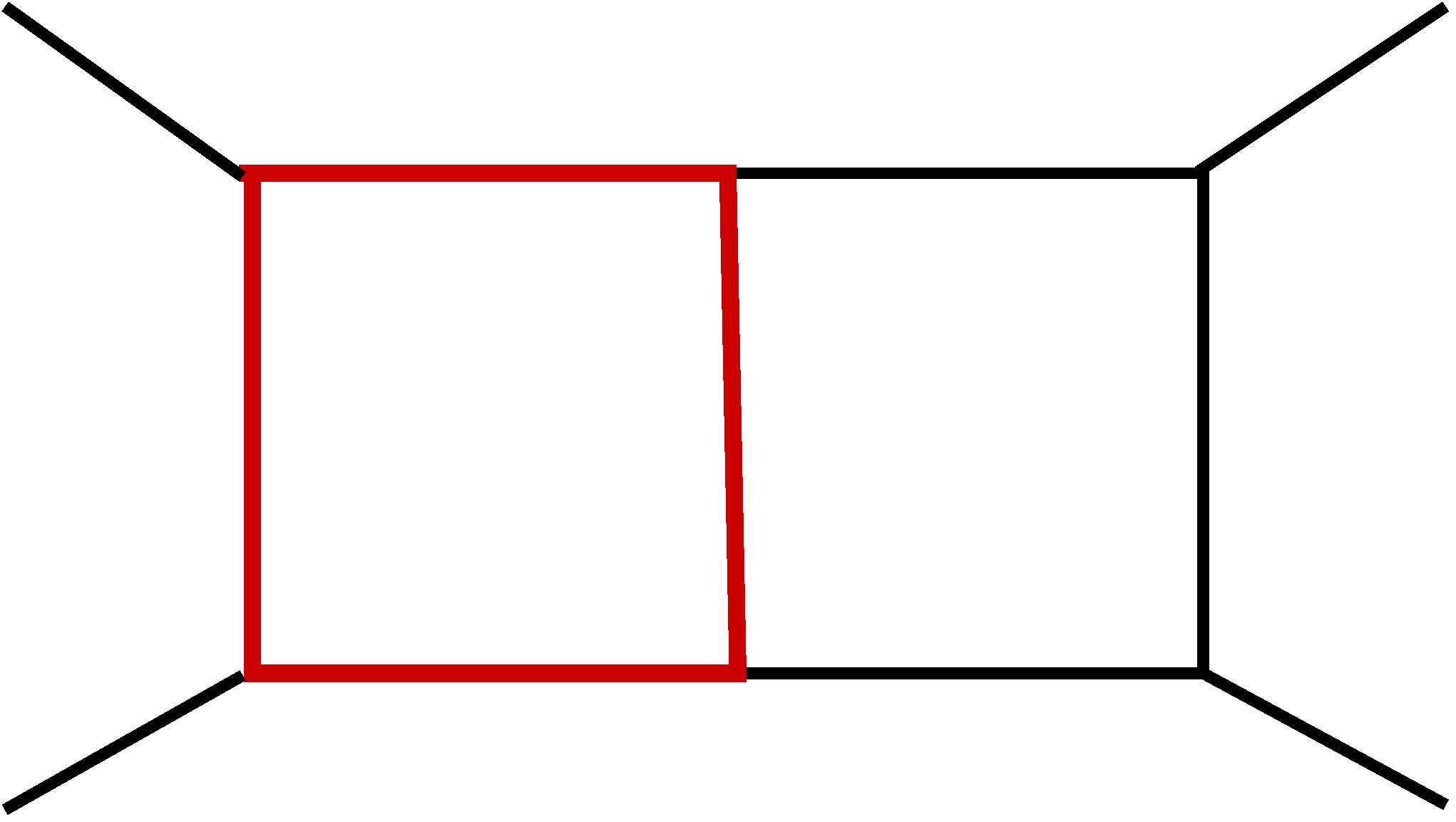}
\hspace{0.5cm}
\includegraphics[width = 1in]{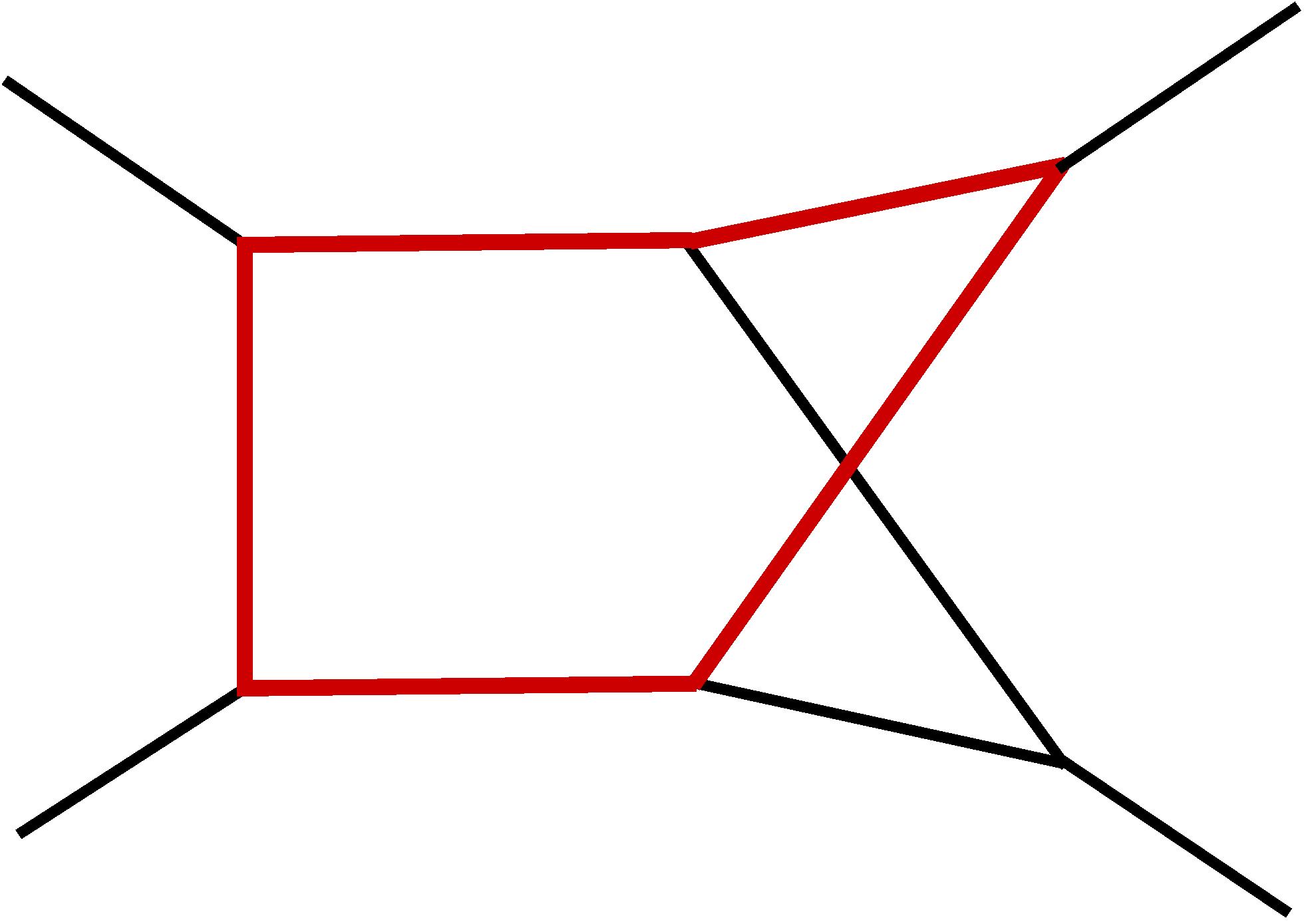}
 \hspace{0.5cm}
\includegraphics[width = 1.1in]{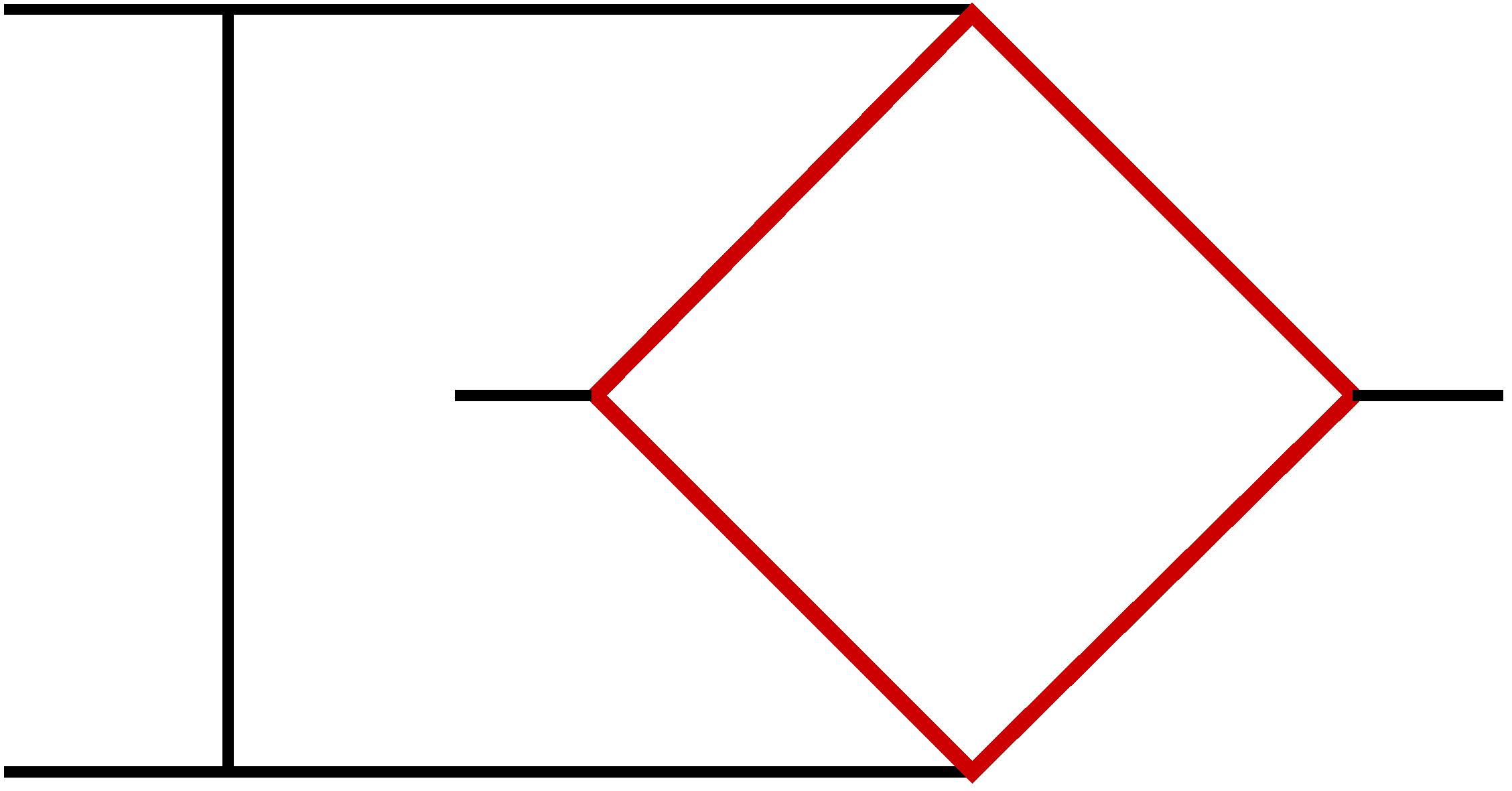}
\caption{{{Topology diagrams respectively for PL1, PL2, NPL1, NPL2 in the top sector.  Red lines represent massive particles, while  black lines denote massless ones.}}}
\label{fig:top-fam-figs}
\end{figure}
We present the top sector diagrams for each integral family in figure~\ref{fig:top-fam-figs}. 

The integrals appearing in the form factors are not all linearly independent. To identify symmetry relations among the integrals, we employ \texttt{Reduze2}~\cite{Studerus:2009ye,vonManteuffel:2012np}. Subsequently, we use \texttt{Kira}~\cite{Maierhofer:2017gsa,Klappert:2020nbg} and  \texttt{LiteRed}~\cite{Lee:2012cn}, which are implementation of the Laporta algorithm~\cite{Laporta:2001dd}, and \texttt{FiniteFlow}~\cite{Peraro:2019svx}, to solve integration-by-parts (IBP) relations. This algorithm leverages finite field arithmetic~\cite{vonManteuffel:2014ixa,vonManteuffel:2016xki,Peraro:2016wsq,Peraro:2019svx} to systematically reduce the integrals to a minimal, independent basis set of master integrals (MIs). Specifically, we obtain 29 MIs in family PL1, 32 in PL2, 54 in NPL1, and 36 in NPL2. Over the past decades, these integrals have been studied in many different contexts~\cite{Caron-Huot:2014lda,Becchetti:2017abb,AH:2023ewe,Becchetti:2023wev,Becchetti:2023yat}. The solution in terms of analytic functions of one of the non-planar topologies involving elliptic sector only became available very recently~\cite{Ahmed:2024tsg} by some of us. This was the last missing piece to achieve two-loop amplitudes in terms of analytic functions. The required master integrals for the amplitudes correspond to the families {PL1, PL2, NPL1 and NPL2} and their crossings: \{$p_1 \leftrightarrow$ $p_2$\}, \{$p_1\rightarrow p_2,p_2 \rightarrow p_3,p_3 \rightarrow p_1$\}, \{$p_1 \rightarrow p_2,p_2 \rightarrow p_4,p_4 \rightarrow p_1$\}, \{$p_1 \rightarrow p_2,p_2 \rightarrow p_3,p_3 \rightarrow p_4,p_4 \rightarrow p_1$\} and \{$p_1 \rightarrow p_2,p_2 \rightarrow p_4,p_4 \rightarrow p_3,p_3 \rightarrow p_1$\}.
 Across all crossings, we find 65 master integrals for the $q\bar{q}$
  channel and 173 for the
$gg$ channel. We also report that while setting up this unified IBP system, we observe that some mappings between integrals are overlooked by the modern IBP softwares. For example, although the system initially yields 91 master integrals, we identify 3 additional relations among them that are not captured.  We derive these missing relations  by mapping the integrals back to the IBP reductions of the individual families.

Although the results of most of the relevant integrals exist in some forms, we set up one system of differential equation containing all master integrals from uncrossed families in order to get the solutions in a consistent representation. The solutions for the crossed integrals are then obtainable by applying the corresponding mapping relations. The integrals presented in ~\cite{Caron-Huot:2014lda,Becchetti:2017abb,AH:2023ewe,Becchetti:2023wev,Becchetti:2023yat,Ahmed:2024tsg} enable a convenient expression of the amplitudes in terms of a canonical basis for all crossings, following the mappings outlined above.  The numerical evaluation of polylogarithmic integrals can be performed in multiple ways. For instance, integrals expressible in terms of Goncharov polylogarithms (MPLs) can be evaluated using GINAC~\cite{Vollinga:2004sn}, while numerically evaluating one-fold integrals over polylogarithmic kernels, also known as dlog one-forms, as done in~\cite{AH:2023ewe} provides another option. Similarly, the numerical evaluation of elliptic kernels can be achieved by series expanding the corresponding kernels along suitable paths in the physical phase-space region, as demonstrated in~\cite{Badger:2021owl,Chaubey:2021ret}. The formulation of these integrals in a function basis suitable for numerical evaluation across all phase-space regions is left for future work.


\section{Ultraviolet and Infrared Structures}
\label{sec:subtraction}
The result of the computation described in the previous section are the divergent helicity amplitudes for the processes described in eq.~\eqref{eq:processes} in terms of bare $\asb$ and bare top mass $m_{tb}$.  
In the following, we describe the ultraviolet (UV) renormalisation and infrared (IR) subtraction of the divergent amplitudes.

\subsection{UV Renormalisation}
\label{sec:uv}
For UV singularity we renormalise the amplitude using the modified minimal subtraction ($\overline{\mathrm{MS}}$) scheme, except for the top quark mass which we choose to renormalise in on-shell (OS) scheme.
The bare coupling $\asb$ is written in terms of the renormalised coupling $\as(\mu)$ as
\begin{align}
\label{bare_to_phys}
\asb \: \mu_0^{2\epsilon} \: S_\epsilon &= \as \: \mu^{2\epsilon} Z_\alpha(\alpha_s(\mu)) ,
\end{align}
where $S_\epsilon = (4 \pi)^{\epsilon} e^{-\gamma_E \epsilon}$, and $\mu$ is the renormalization scale, which we set equal to $\mu_0$. The latter is introduced in dimensional regularisation to make the coupling constant dimensionless. 
The bare top-quark mass, $m_{t,b}$, is expressed in terms of the renormalized mass, $m_t$, as:
\begin{align}
\label{bare_to_physmass}
m_{t,b}  S_\epsilon = m_t Z_{m_t},
\end{align}
where $Z_{m_t}$ is the mass renormalization constant.
Similarly, the bare gluon field, ${\mathcal{G}_{\nu, b}}$, is related to the renormalized gluon field, ${\mathcal{G}_\nu}$, via:
\begin{align}
\label{bare_to_physG}
{\mathcal{G}_{\nu, b}}  S_\epsilon = {\mathcal{G}_\nu} Z_g,
\end{align}
where $Z_g$ is the gluon field renormalization constant. This arises due to the presence of massive quark. 
The bare quark field, $Q_b$, is connected to the renormalized one, $Q$, as:
\begin{align}
\label{bare_to_physq}
{Q}_{b}  S_\epsilon = {Q} Z_q,
\end{align}
where $Z_q$ represents the quark field renormalization constant.
We set $n_f = 0$ as there are no massless quark loops contributing to the processes described in eq.~\eqref{eq:processes} at the perturbative order considered.

\subsubsection*{Gluon Channel}
\begin{figure}[htb]
\centering
\includegraphics[width = 1.1in]{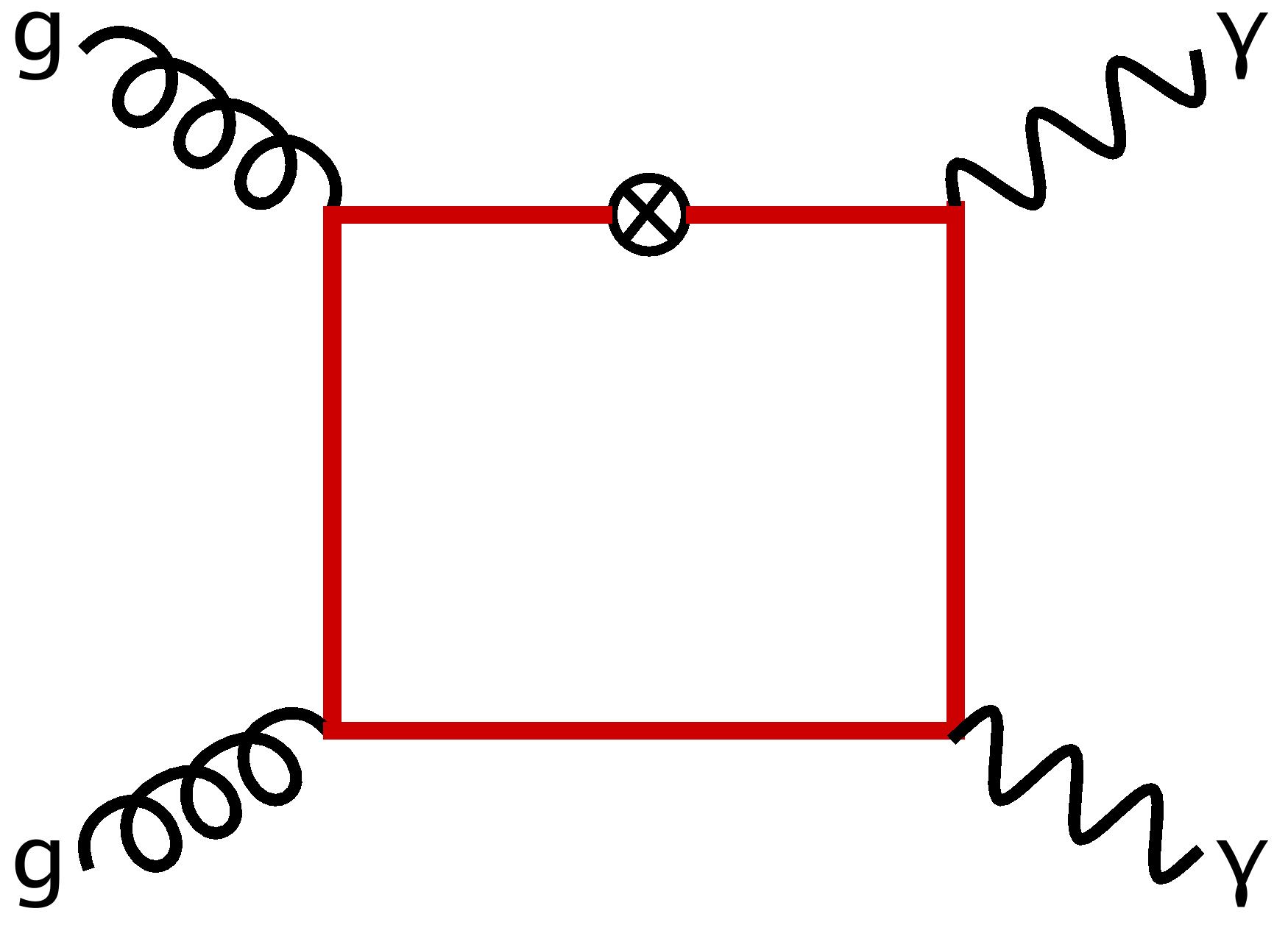}
\caption{Sample diagram for calculating mass counter term.}
\label{fig:ggphph-mCT}
\end{figure}

Since the leading-order $g g \to \gamma \gamma$ amplitude is loop-induced, as shown in figure~\ref{fig:ggaa-qqaa-1L}, it is free from both ultraviolet (UV) and infrared (IR) divergences. At two-loop level, however, the amplitude exhibits both UV and IR divergences. Notably, only a single massive quark loop contributes to the amplitude - photons can only emit from massive quarks. In other words, the two-loop amplitude does not depend on massless quarks. Therefore, we can safely disregard the massless quark contributions from the leading order when constructing the UV and IR subtraction terms.
The UV renormalized helicity amplitude, $\mathcal{H}_{\vec{\lambda},\text{ren}}^{g,(l)}$ with $l = 1, 2$, is obtained from the bare helicity amplitude defined in eq.~\eqref{eq:helicity_pert} using the following:
\begin{align}
\mathcal{H}_{\vec{\lambda},\text{ren}}^{g,(1)} &=  \mathcal{H}_{\vec{\lambda}}^{g,(1)} , \nonumber\\
\mathcal{H}_{\vec{\lambda},\: \text{ren}}^{g,(2)} &= \mathcal{H}_{\vec{\lambda}}^{g,(2)} + \left(\frac{n_g}{2} \delta Z_g + \delta Z_\alpha\right)  {\mathcal{H}^{g,(1)}_{\vec{\lambda},\text{ren}}} + \delta Z_m \;
 \mathcal{H}^{g,CT,(1)}_{\vec{\lambda}}.
\label{hel_amp_ren}
\end{align}
Here the renormalisation constants are expanded according to $Z_i = 1 + \as \delta Z_i + 
 \mathcal{O}(\as^2)$ for $i= \alpha, g, m$ with
 \begin{align}
 \delta Z_\alpha &=-\frac{\beta_0 }{\epsilon} + {\left(\frac{\mu^2}{{m_t}^2}\right)}^{\ep} \left(\frac{4}{3\ep}T_F\right) n_{f_t} \,,\nonumber\\
 \delta Z_g &= - \left(\frac{\mu^2}{{m_t}^2}\right)^\ep\left(\frac{4}{3\ep}T_F \right) n_{f_t} \,,\nonumber\\
 \delta Z_m &= - \left(\frac{\mu^2}{{m_t}^2}\right)^\ep C_F \left(\frac{3}{\ep} + 4 \right)\,. 
 \label{eq:renormconst}
 \end{align}
The number of gluons in the external states is denoted by $n_g$ which is equal to 2 in our case. The quadratic Casimir in the fundamental representation of SU($N$) is $C_F = (N^2 - 1) / (2N)$, and in the adjoint representation, it is denoted by $C_A$. The constant $T_F$ is defined as $T_F = 1/2$, and the leading-order $\beta$ function is given by $\beta_0 = (11 C_A - 2 n_f) / 3$. It is noteworthy that the top-mass-dependent contributions to the $\alpha_s$ expansions from $\delta Z_\alpha$ and $\delta Z_g$ cancel each other. The counter-term amplitude for the top mass renormalization is represented by $\mathcal{H}^{g,CT,(1)}_{\vec{\lambda}}$. This counter-term amplitude is derived by inserting the mass counter-term, $\mathcal{P}^{m_t}_{ac}$, defined through 
\begin{align}
    \mathcal{P}^{m_t}_{ac} = \frac{\textit{i}\; \delta_{ab}}{\cancel{p} - m_t} \left(- \textit{i}\;\delta Z_m\right) \frac{\textit{i} \;\delta_{bc}}{\cancel{p} - m_t},
\label{eq:ct}
\end{align}
into each top quark propagator in the leading-order amplitude, and collecting the coefficient of $\as^2$. This can be visualised through figure~\ref{fig:ggphph-mCT}.
Alternatively, the counter-term can be computed by differentiating the leading-order amplitude with respect to $m_{t}$. This approach yields results that are in perfect agreement with the previously derived counter-term.

\subsubsection*{Quark Channel}
In the quark-initiated channel, one-loop diagrams containing a single massive quark loop exist but vanish due to Furry's theorem. Non-zero contributions begin to appear only at the two-loop level. These contributions can be categorized into two types of diagrams, depending on whether the photons are emitted from massive or massless quarks, as illustrated in figure~\ref{fig:process1-figs}.
The first type, where photons are emitted from massive quarks, is UV and IR finite. This behaviour is expected since no such diagrams exist at lower loop levels. The second type, involving photons emitted from massless quarks, is UV divergent but IR finite. Calculating the counterterms requires considering tree-level and one-loop diagrams without massive quark involvement.
Thus, while we focus on diagrams with at least one massive quark loop, the two-loop UV and IR subtraction contributions also include contributions from massless quarks which are not forming closed loops.

We split the helicity amplitude defined in eq.~\eqref{eq:helicity_pert} with respect to the type of quarks from which the di-photon are emitted:
\begin{align} 
\label{eq:quarkchnl}
\mathcal{H}^{q,(\ell)}_{\vec{\lambda}} &= {Q^{2}_f}\;{\mathcal{H}^{f,(\ell)}_{\vec{\lambda}}} + \sum_{f_t=1}^{n_{f_t}}Q^{2}_{f_t}\;{\mathcal{H}^{f_t,(\ell)}_{\vec{\lambda}}}.
\end{align}
The terms ${\mathcal{H}^{f,(\ell)}_{\vec{\lambda}}}$ and ${\mathcal{H}^{f_t,(\ell)}_{\vec{\lambda}}}$ represent the contributions from diagrams where the diphoton is emitted by massless and massive quarks, respectively. Notably, there are no non-zero mixed diagrams up to two loops.
As previously mentioned, the contribution ${\mathcal{H}^{f_t,(2)}_{\vec{\lambda}}}$ is both UV and IR finite, because it first arises at the two-loop level.
On the other hand, the contribution ${\mathcal{H}^{f,(2)}_{\vec{\lambda}}}$ is UV divergent but remains IR finite. In QCD, $n_{f_t}$ equals 1.

Additionally, we require massless quark field renormalisation constants up to order $\as^2$ along with the constants of eq.~\eqref{eq:renormconst}:
\begin{align}
Z_q = 1 + \left(\frac{\as}{4\pi}\right) \delta {Z_q}^{(1)}+  \left(\frac{\as}{4\pi}\right)^2 \delta Z_q^{(2)} + \mathcal{O}(\as^3),
\end{align}
with
\begin{align}
    &\delta Z_q^{(1)} = 0,\nonumber\\
    &\delta Z_q^{(2)} = {\left(\frac{\mu^2}{{m_t}^2}\right)}^{\ep} C_F \; n_{f_t} \left(\frac{1}{4 \ep} - \frac{5}{4} \right).
\end{align}
We need to consider only ${\mathcal{H}^{f,(\ell)}_{\vec{\lambda}}}$ for renormalisation and we obtain the UV finite amplitude ${\mathcal{H}^{f,(2)}_{\vec{\lambda}, \text{ren}}}$ by
\begin{align}
{\mathcal{H}^{f,(2)}_{\vec{\lambda},\text{ren}}} &= {\mathcal{H}^{f,(2)}_{\vec{\lambda}}} + \delta {Z_q}^{(1)} {\mathcal{H}^{f,(1)}_{\vec{\lambda}}} + \delta{Z_\alpha}{\mathcal{H}^{f,(1)}_{\vec{\lambda}}} + \delta Z_q^{(2)} \mathcal{H}_{\vec{\lambda}}^{f,(0)}.
\label{hel_amp_renquark}
\end{align}
$\mathcal{H}_{\vec{\lambda}}^{f,(0)}$ is tree level and $\mathcal{H}_{\vec{\lambda}}^{f,(1)}$ is the one loop helicity amplitude  setting $n_f$ to zero in $q \bar{q} \to \gamma \gamma$ channel, respectively.

\subsection{IR Factorisation}
The IR singularity structure of QCD amplitudes has been studied up to three loops for the massless cases in refs.~\cite{Catani:1998bh,Sterman:2002qn,Aybat:2006wq,Aybat:2006mz,Becher:2009cu,Becher:2009qa,Dixon:2009gx,Gardi:2009qi,Gardi:2009zv,Almelid:2015jia}. It also has been extended to the cases involving massive partons at two loops in refs.~\cite{Catani:2000ef,Ferroglia:2009ep,Ferroglia:2009ii,Mitov:2009sv,Mitov:2010xw} and up to three loops~\cite{Liu:2022elt} involving one massive parton in the external states.
The IR divergences can be subtracted
from our UV renormalized amplitudes, $\mathcal{H}_{\vec{\lambda},\:\text{ren}}$, multiplicatively through
%
%
\begin{equation}
\label{eq:IR_factorization}
\mathcal{H}_{\vec{\lambda},\:\text{fin}}^{g|q} =\lim_{\epsilon \to 0} \left[ \mathcal{Z}^{-1}_{\rm IR} \mathcal{H}_{\vec{\lambda},\:\text{ren}}^{g|q} \right]_{ \as^{QCD}\to\xi\as},
\end{equation}
resulting IR finite $\mathcal{H}_{\vec{\lambda},\:\text{fin}}^{g|q}$.
Here $\as$ denotes the strong coupling constant in the effective theory with $n_f=5$  in which the heavy quark is integrated out. While considering an amplitude with heavy quark mass dependence, one must relate the $\as^{QCD}$, the strong coupling constant of full QCD with $n_f=6$ through the decoupling relation~\cite{Steinhauser:2002rq}, $\as^{QCD}\;=\;\xi\as$. Where the $\xi$ to the order of $\as$ is given by 
 \begin{equation}
     \xi = 1 + \frac{\as}{4\pi} \;\sum_{i=1}^{n_{f_t}} \frac{2}{3} \left[\; e^{\ep\gamma_{E}}\; \Gamma(\ep) \;\left(\frac{\mu^2}{m_i^2}\right)^{\ep}  - \frac{1}{\ep} \right]\,.
     \end{equation}
Here, $\mathcal{Z}_{\rm IR}$ is a matrix in SU(N) color space acting on the space spanned by the $\mathcal C_i$ basis vectors \eqref{eq:color} and $\mathcal{H}_{\vec\lambda,\:\text{fin}}^{g|q}$ are finite remainders, also called hard scattering functions.
The matrix $\mathcal{Z}_{\rm IR}$ can be written as 
\begin{equation}\label{exponentiation_B}
\mathcal{Z}_{\rm IR} = \mathbb{P} \exp \left[
  \int_\mu^\infty \frac{\mathrm{d} \mu'}{\mu'}
  \mathbf{\Gamma}(\{p\},\{m_t\},\mu')\right], 
\end{equation}
where $\mathbb{P}$ denotes the path-ordering
of color operators~\cite{Becher:2009qa} in increasing values of $\mu'$ from left to right.
The anomalous dimension matrix $\mathbf{\Gamma}=$ $\mathbf{\Gamma}_{\text{dipole}}$ can be written as
\begin{equation}\label{eq:dipole}
\mathbf{\Gamma}_{\text{dipole}}(\{p\},\mu)  =  \sum_{1\leq i < j \leq 2} \mathbf{T}_i \cdot  \mathbf{T}_j  \; \gamma^{\text{K}}(\as) \; \log\left(\frac{\mu^2}{-s_{ij}-i\delta}\right) + \sum_{i=1}^2 \;  \gamma^i(\as)\; , 
\end{equation}
where $\gamma^{\text{K}}(\as)$ is the \emph{cusp anomalous
dimension} \cite{Korchemsky:1987wg,Moch:2004pa,Vogt:2004mw,Bruser:2019auj,Henn:2019swt,vonManteuffel:2020vjv} and $\gamma^i$ the quark (gluon) \emph{collinear anomalous dimension} \cite{Ravindran:2004mb,Moch:2005id,Moch:2005tm,Agarwal:2021zft} of the $i$-th
external particle.
Further, $\mathbf{T}^a_i$ represents the color generator of the $i$-th parton in the scattering amplitude,
\begin{alignat}{2}\label{convention}
(\mathbf{T}^a_i)_{\alpha\beta} &=  t^a_{\alpha \beta} \; &&\text{ for a final(initial)-state quark (anti-quark)}, \nonumber\\
(\mathbf{T}^a_i)_{\alpha\beta} &=  -t^a_{\beta\alpha} \; &&\text{ for a final(initial)-state anti-quark (quark)}, \nonumber\\
(\mathbf{T}^a_i)_{bc} &=  -if^{abc} \;
&&\text{ for a gluon.}
\end{alignat}
As the processes, we considered in ~\eqref{eq:processes} do not have any massive parton in the external states, we exclude the contributions from massive parton in the external states in ~\eqref{eq:dipole}.
We expand the finite remainders in powers of $\alpha_s$ as 
\begin{equation}
\mathcal{H}_{\vec\lambda,\text{fin}}^{g|q}=\sum_{\ell\geq 0}\as^\ell \mathcal{H}^{{g|q},(\ell)}_{\vec\lambda,\text{fin}}\,.
\end{equation}
As previously mentioned, the $q\bar q \to \gamma\gamma$ does not exhibit any IR divergences. We need IR subtraction only for the $gg \to \gamma\gamma$ channel. The finite remainders for the quark-initiated channel are denoted through $\mathcal{H}_{\vec\lambda,\:\text{fin}}^{f,(l)}$ and $\mathcal{H}_{\vec\lambda,\:\text{fin}}^{f_t,(l)}$. For the gluon-initiated channel, the corresponding expressions are given by 
\begin{align}
\label{one}
&\mathcal{H}_{\vec\lambda,\:\text{fin}}^{g,(0)} = \mathcal{H}_{\vec\lambda}^{g,(0)} \;,\nonumber\\
&\mathcal{H}_{\vec\lambda,\:\text{fin}}^{g,(1)} = \mathcal{H}_{\vec\lambda,\: \text{ren}}^{g,(1)} - \mathcal{Z}^{(1)}_{\rm IR}\mathcal{H}_{\vec\lambda,\: \text{ren}}^{g,(0)},
\end{align}
where $\mathcal{Z}^{(n)}_{\rm IR}$ are the coefficients of the expansion of $\mathcal{Z}_{\rm IR}$ in $\as$~\cite{Becher:2009qa, Caola:2021rqz}:
\begin{equation}\label{eq:Zir}
 \mathcal{Z}^{(0)}_{\rm IR} =  1 \,,\nonumber\\
 \mathcal{Z}^{(1)}_{\rm IR} =  \frac{\Gamma'_0}{4 \ep^2} + \frac{\mathbf{\Gamma}_0}{2 \ep}\;.
 \end{equation}
 The quantity $\Gamma'(\as)$ is defined through
\begin{equation}
\Gamma'(\as) = \frac{\partial \mathbf{\Gamma}(\{p\},\as,\mu)}{\partial \log \mu} =  -\gamma^{\text{K}} \sum_i C_i  = \sum_{\ell \geq 0} \as^{\ell+1} \Gamma_\ell' ,
\end{equation}
with the last equal sign defining the perturbative coefficients $\Gamma_\ell' $.


\section{Results, Checks and Benchmarks}
\label{sec:res}
Upon including the UV counterterms, we confirm the complete cancellation of UV divergences. While not all amplitudes under consideration exhibit IR divergences, for those that do, the soft and collinear singularities align precisely with theoretical predictions, as described in section~\ref{sec:subtraction}. This consistency is reflected in the finiteness of $\mathcal{H}_{\vec{\lambda},:\text{fin}}^{g|q}$ in eq.~\eqref{eq:IR_factorization}. This agreement serves as a crucial validation of our calculation. 
An independent calculation of the helicity amplitudes is carried out in ref.~\cite{Becchetti:2025diph}, and we find perfect numerical agreements with their bare results.

As previously mentioned, due to the lack of a suitable functional basis for numerical evaluation at the time of publication, we use AMFlow~\cite{Liu:2022chg} to calculate finite remainders numerically at some kinematic points. To systematically represent these results, we parameterise the physical kinematic space as~\cite{Becchetti:2023wev}.
\begin{align}
    s > 0, t=-\frac{s}{2} (1-cos\; \theta), -s <t<0.
\end{align}
The scattering angle in the partonic centre of mass frame is denoted by $\theta \in (0,\pi)$. 
Table~\ref{tab:benchmark-quark} and~\ref{tab:benchmark-gluon} provide benchmark values for the two-loop finite remainders of all helicity amplitudes at selected kinematic points in the physical phase space. 
\begin{table}[t!]
    \centering
    \renewcommand{\arraystretch}{1.5} 
    \setlength{\tabcolsep}{10pt} 
    \begin{tabular}{|c|c|}
        \hline
       Helicity  &   {Finite remainder} \\
        \hline
        ${\mathcal{H}^{f,(2)}_{-+--,{\rm fin}}}$   &0.0003077743812 \\
        ${\mathcal{H}^{f,(2)}_{-+-+,{\rm fin}}}$ &  0.3343545753 + 0.1052222825 I \\
        ${\mathcal{H}^{f,(2)}_{-++-,{\rm fin}}}$ &  -7.657428648 + 
  6.311781761 I \\
        ${\mathcal{H}^{f,(2)}_{-+++,{\rm fin}}}$ & -0.0003077743812 \\
        \hline
        ${\mathcal{H}^{f_t,(2)}_{-+--,{\rm fin}}}$  &0.0004586738171 I\\
        ${\mathcal{H}^{f_t,(2)}_{-+-+,{\rm fin}}}$    & - 0.08920732713 I \\
        ${\mathcal{H}^{f_t,(2)}_{-++-,{\rm fin}}}$    & 0.006847175765 I \\
        ${\mathcal{H}^{f_t,(2)}_{-+++,{\rm fin}}}$   & - 0.0004586738171 I  \\
        \hline
    \end{tabular}
    \caption{Benchmarks for the finite remainders for the quark channel for $\theta=\frac{\pi}{6}$, $s= 3$ GeV and $N=3$. }
    \label{tab:benchmark-quark}
\end{table}
\begin{table}[h!]
    \centering
    \renewcommand{\arraystretch}{1.5} 
    \setlength{\tabcolsep}{10pt} 
    \begin{tabular}{|c|c|c|c|}
        \hline
       Helicity  &   {Finite remainder} \\
        \hline
        ${\mathcal{H}^{g,(2)}_{++++,{\rm fin}}}$   & -4.235 + 44.746 I \\
        ${\mathcal{H}^{g,(2)}_{-+++,{\rm fin}}}$ &  -0.21954 + 0.61617 I \\
        ${\mathcal{H}^{g,(2)}_{+-++,{\rm fin}}}$ &  -0.21954 + 0.61617 I \\
        ${\mathcal{H}^{g,(2)}_{++-+,{\rm fin}}}$ &  -0.35594 + 0.22179 I \\
        ${\mathcal{H}^{g,(2)}_{+++-,{\rm fin}}}$ & -0.35594 + 0.22179 I  \\
        ${\mathcal{H}^{g,(2)}_{--++,{\rm fin}}}$ &  0.920 - 69.801 I \\
        ${\mathcal{H}^{g,(2)}_{-+-+,{\rm fin}}}$ & 67.226 - 69.286 I   \\
        ${\mathcal{H}^{g,(2)}_{+--+,{\rm fin}}}$ & 0.40969 - 0.44848 I \\        
        \hline
    \end{tabular}
  
    \caption{Benchmarks for the finite remainders for the gluonic channel for $\theta=\frac{\pi}{6}$, $s= 3$ GeV and $N=3$.}
    \label{tab:benchmark-gluon}
\end{table}

The existence of Bose symmetry due to the exchange of final state photons, $p_3 \leftrightarrow p_4$ is evident from the finite remainder for both quark and gluon-initiated processes. For the $q\bar q$ channel, it gets translated to
\begin{align}
    &\mathcal{H}^{f|f_t,(2)}_{-+--,\rm fin}(s,t) = -\mathcal{H}^{f|f_t,(2)}_{-+++,\rm fin}(s,t),\nonumber\\
    &\mathcal{H}^{f|f_t,(2)}_{-+-+,\rm fin}(s,t) = \mathcal{H}^{f|f_t,(2)}_{-++-,\rm fin}(s,u).
\end{align}
The notation $f|f_t$ signifies that the relations hold for both types of finite remainders. These validations serve as crucial consistency checks on our final results.
For the gluon-initiated amplitude, the Bose-symmetry under the exchange of $p_1 \leftrightarrow p_2$ and/or $p_3 \leftrightarrow p_4$ implies
\begin{align}
    &\mathcal{H}^{g,(2)}_{\lambda_1,\lambda_2,\lambda_3,\lambda_4,\rm fin}(s,t) = \mathcal{H}^{g,(2)}_{\lambda_2,\lambda_1,\lambda_3,\lambda_4,\rm fin}(s,u)\,\nonumber\\
    &\mathcal{H}^{g,(2)}_{\lambda_1,\lambda_2,\lambda_3,\lambda_4,\rm fin}(s,t) = \mathcal{H}^{g,(2)}_{\lambda_1,\lambda_2,\lambda_4,\lambda_3,\rm fin}(s,u)\,.
\end{align}
The finite remainders are checked to exhibit this symmetry.
We provide the bare helicity amplitudes expressed in terms of a set of master integrals as an ancillary file~\cite{zenodokaur25}. The finite remainder is available upon request from the authors.

\section{Conclusions}
\label{sec:conc}

We compute the two-loop QCD helicity amplitudes for $gg \to \gamma\gamma$ and $q\bar{q} \to \gamma\gamma$, retaining the full dependence on the top quark mass inside the loop. Using a combination of in-house and publicly available codes, we express the integrand in terms of a set of master integrals. A recent computation by some of us~\cite{Ahmed:2024tsg} involving a non-planar integral family with elliptic sectors provides the final missing ingredient, allowing us to complete this calculation. While the remaining required master integrals exist in the literature, we perform an independent validation by constructing a comprehensive system of differential equations encompassing all master integrals (modulo crossings). This ensures a consistent representation of the solutions in terms of a unified set of variables.  This set of uncrossed families and the corresponding function basis remain the same for dijet production. Therefore, while we defer the publication of these results to future work, we provide the bare amplitudes in terms of a chosen set of master integrals as an ancillary file~\cite{zenodokaur25} with this article.

We renormalise the heavy quark mass in the on-shell scheme, while other quantities are renormalised in the $\overline{\text{MS}}$ scheme. In addition to verifying the expected UV and IR divergences, we cross-check our bare amplitudes with an independent calculation by another group~\cite{Becchetti:2025diph}, finding complete numerical agreement at multiple physical phase-space points. We present a few benchmark values for the finite remainders for all helicity amplitudes. 

These amplitudes provide the foundation for computing cross-sections and other key observables using various subtraction schemes. It will be interesting to investigate the impact of these analytic results by comparing them with existing calculations of the diphoton production cross-section for $gg \to \gamma\gamma$, where the relevant integrals were previously evaluated numerically~\cite{Maltoni:2018zvp} or semi-numerically~\cite{Chen:2019fla}. 
The impact of the top quark mass at the high-luminosity phase of the LHC will be particularly interesting to explore, as its effects are expected to be significantly enhanced in this regime. Furthermore, this work lays the groundwork for future studies, including dijet production with a massive quark loop.  
\section*{Acknowledgments}
TA and AC is grateful to Raoul Röntsch and Vajravelu Ravindran for useful discussions. AC is also indebted to Andreas von Manteuffel for the help regarding the generation of the shift identities in \texttt{Reduze2}~\cite{Studerus:2009ye,vonManteuffel:2012np}. AC has been
supported by the Italian Ministry of Universities and
Research through Grant No. PRIN 2022BCXSW9. The work of EC is funded by the ERC grant
101043686 ‘LoCoMotive’.
\bibliographystyle{JHEP}
\bibliography{main} 
\end{document}